\documentclass[10pt,aps,preprintnumbers,floats,prd,nofootinbib]{revtex4-1}

\usepackage{graphicx}
\usepackage{dcolumn}
\usepackage{amsmath}
\usepackage{amsthm}

\usepackage{amsfonts}
\usepackage{euscript,bbm}
\usepackage{ifthen}
\usepackage{psfrag}
\usepackage{slashed}

\pdfoutput=1
\usepackage{graphicx}
\usepackage{cancel}
\usepackage{amssymb}
\usepackage{amsmath,dsfont,textcomp}
\usepackage{color}
\usepackage{euscript,amsmath}
\usepackage{epsf}
\usepackage{hyperref}
\usepackage{multirow}

\usepackage{subfigure}
\usepackage{slashed}
\usepackage{simplewick}

\def\lsim{\mathrel{\rlap{\lower4pt\hbox{\hskip1pt$\sim$}}
    \raise1pt\hbox{$<$}}}         
\def\gsim{\mathrel{\rlap{\lower4pt\hbox{\hskip1pt$\sim$}}
    \raise1pt\hbox{$>$}}}

\def\beq{\begin{equation}}
\def\eeq{\end{equation}}
\def\bea{\begin{eqnarray}}
\def\eea{\end{eqnarray}}
\def\bitem{\begin{itemize}}
\def\eitem{\end{itemize}}
\def\ba{\begin{array}}
\def\ea{\end{array}}
\def\bal{\begin{align}}
\def\eal{\end{align}}
\def\bi{\begin{itemize}}
\def\ei{\end{itemize}}
\def\lsim{\mathrel{\rlap{\lower4pt\hbox{\hskip1pt$\sim$}}
    \raise1pt\hbox{$<$}}}         
\def\gsim{\mathrel{\rlap{\lower4pt\hbox{\hskip1pt$\sim$}}
    \raise1pt\hbox{$>$}}}

\newcommand{\newc}{\newcommand}
\newc{\renewc}{\renewcommand}

\newc{\ie}{{\it i.e.~}}          \newc{\etal}{{\it et al.~}}
\newc{\eg}{{\it e.g.~}}          \newc{\etc}{{\it etc.~}}
\newc{\cf}{{\it c.f.~}}
\newc{\os}{\mbox{\hspace{4pt}}}
\newc{\us}{\mbox{\hspace{12pt}}}

\renewc{\bar}{\overline}

\newc{\gev}{\,{\rm GeV}}
\newc{\mev}{\,{\rm MeV}}
\newc{\ev}{\,{\rm eV}}
\newc{\kev}{\,{\rm keV}}
\newc{\tev}{\,{\rm TeV}}
\def\ln{\mathop{\rm ln}}

\def\Tr{\mathop{\rm Tr}}

\newc{\LM}{\mathcal{L}}
\newc{\SM}{\mathcal{S}}

\newc{\HM}{\mathcal{H}}
\newc{\GM}{\mathcal{G}}
\newc{\OM}{\mathcal{O}}
\newc{\FM}{\mathcal{F}}
\newc{\AM}{\mathcal{A}}
\newc{\BM}{\mathcal{B}}
\newc{\NM}{\mathcal{N}}
\newc{\WM}{\mathcal{W}}
\newc{\ZM}{\mathcal{Z}}

\newc{\Chi}{\mathcal{X}}

\newcommand{\be}{\begin{equation}}
\newcommand{\ee}{\end{equation}}

\newcommand{\nn}{\nonumber}

\newcommand{\braket}[1]{\left<#1\right>}

\newcommand{\met} {{E\!\!\!\!/_{\rm T}}}

\newcommand{ \pgs }    {{\tt PGS4}}

\begin{document}

\begin{flushright}
MIFPA-14-11
\end{flushright}


\title{\bf\Large Exploring the Doubly Charged Higgs of the Left-Right Symmetric Model using Vector Boson Fusion-like Events at the LHC }


\author{Bhaskar Dutta$^{1}$, Ricardo Eusebi$^1$, Yu Gao$^{1}$, Tathagata Ghosh$^{1}$ and Teruki Kamon$^{1,2}$}

\affiliation{$^{1}$~Mitchell Institute for Fundamental Physics and Astronomy, \\
Department of Physics and Astronomy, Texas A\&M University, College Station, TX 77843-4242, USA \\
$^{2}$~Department of Physics, Kyungpook National University, Daegu 702-701, South Korea 
}

\begin{abstract}

This paper studies the pair production of the doubly charged Higgs boson of the left-right symmetric models using multilepton final state in the vector boson fusion (VBF)-like processes. The study is performed in the framework consistent with the model's correction to the standard model $\rho_{EW}$ parameter. VBF topological cuts, number of leptons in the final state and $p_T$ cuts on the leptons are found to be effective in suppressing the background. Significant mass reach can be achieved for exclusion/discovery of the doubly charge Higgs boson for the upcoming LHC run with a luminosity of $\mathcal{O}(10^3)$ fb$^{-1}$.
\end{abstract}
\maketitle

\section{Introduction}

The LHC experiments have successfully discovered the last missing piece of the standard model (SM) - the elusive Higgs boson. However no sign of any physics beyond the SM has been observed yet. Although the standard model has been extremely effective, many unsolved questions still remain. The left-right (L-R) symmetric models \cite{Pati:1974yy,LRModel1,LRModel2,Mohapatra:1979ia} provide appealing solutions for some of these questions. Firstly it explains the origin of the parity violation at the weak scale. In L-R models, parity is an exact symmetry of the weak-interaction Lagrangian at an energy scale much higher than the SM scale. The parity violation arises from the spontaneous symmetry breaking driven by a vacuum not being invariant under parity. Secondly, the L-R models predicts the existence of right handed neutrinos that explain the light neutrino mass via the see-saw mechanism.
Finally these models place quarks and leptons on the same footing in the weak interactions and provide a simple formula for the electric charge, involving only the weak isospin and the difference between the baryon and lepton numbers ($B$-$L$).

In this paper, we explore the LHC signature of these L-R models by focusing on the like-sign leptons from the decays of the left-handed doubly charged scalars ($\delta^{\pm \pm}_{L}$). It is worthwhile to mention here, that the doubly charged Higgs particles may also arise in other models, such as,  Georgi-Machacek model \cite{Georgi:1985nv}, Littlest Higgs model \cite{ArkaniHamed:2002qy} and 3-3-1 models \cite{Frampton:1992wt,Pisano:1991ee}. Recent publications \cite{delAguila:2013mia,delAguila:2013hla,Rentala:2011mr,delAguila:2013yaa} have studied the doubly charged Higgs belonging to different multiplets. A rather interesting work~\cite{Quintero:2012jy} has investigated lepton number violating decays of heavy flavor fermions ($t,\tau$) induced by the doubly charged Higgs.

In Refs. \cite{Han:2007bk,Kanemura:2013vxa,Bambhaniya:2013wza,Akeroyd:2009hb,Cagil:2012py,
Akeroyd:2010ip,Tonasse:2012nq,delAguila:2013mia,Akeroyd:2011zza,Aoki:2011pz,
Chun:2012zu,Sugiyama:2012yw,Azuelos:2004mwa} the production of doubly charged Higgs boson has been considered using Drell-Yan (DY) mechanism in the scope of the models described above. In this paper, we consider the production of a pair of doubly charged Higgs boson accompanied by two energetic tagging jets, predominantly produced by vector boson fusion (VBF) processes, where the leading jets are very helpful to reduce the background. In various different contexts the single production of a doubly charged Higgs boson through $W^{\pm} W^{\pm}$ fusion at the LHC, have been discussed in some earlier studies\cite{Chiang:2012dk,Englert:2013wga,Godfrey:2010qb,Grinstein:2013fia,
Azuelos:2004mwa}. In these studies the vacuum expectation value (VEV) of the triplet sector is assumed to be non-zero and consequently the final states with $W^{\pm} W^{\pm}$ are  considered using the decay modes of $W$ containing $e$ and $\mu$. However we differ from these searches in the sense that, we have only considered the scenario where the left triplet VEV of L-R symmetric models are zero, to avoid unnatural experimental consequences. In such a scenario, we will consider e, $\mu$ and $\tau$ (hadronic) final states arising from the doubly charged Higgs decays directly along with two high $p_T$ jets. By using VBF topological cuts, we shall offer a  search strategy of the doubly charged Higgs boson at the LHC, complementary to the current search being performed by CMS \cite{Chatrchyan:2012ya} and ATLAS \cite{Aad:2012cg,Aad:2012xsa} for multilepton final states  without using two tagged jets. The potential discovery of $\delta^{\pm \pm}_L$ in both VBF and DY channel can be instrumental in determining the electroweak (EW) origin of it.

The paper is orgainized as follows: Section~\ref{L-R Model} briefly reviews the L-R symmetric model. Section~\ref{Pheno} discusses the production mechanism and the subsequent decay of the doubly charged Higgs. The results for future LHC searches are given in Section~\ref{Results}. Finally the conclusions drawn from the study have been summarized in Section \ref{Conclusion}.

\section{The Left-Right Symmetric Model}
\label{L-R Model}

Here we  present a brief overview of the $SU(2)_L \times SU(2)_R \times U(1)_{B-L} $ L-R symmetric models~\cite{Pati:1974yy,LRModel1,LRModel2,Mohapatra:1979ia}. The scalar sector of these models consist of the following Higgs multiplets~\cite{Mohapatra:1980yp,Lim:1981kv,Olness:1985bg,Gunion,Desh}:
 \begin{eqnarray} \phi = \begin{pmatrix}
\phi^{0}_{1} & \phi^{+}_1 \\ 
\phi^{-}_2 & \phi^{0}_2
\end{pmatrix},\end{eqnarray}
\begin{eqnarray} \Delta_L = \begin{pmatrix}
\delta^{+}_L/\sqrt{2} & \delta^{++}_L \\ 
\delta^{0}_L & -\delta^{+}_L/\sqrt{2}
\end{pmatrix},\end{eqnarray} 
\begin{eqnarray} \Delta_R = \begin{pmatrix}
\delta^{+}_R/\sqrt{2} & \delta^{++}_R \\ 
\delta^{0}_R & -\delta^{+}_R/\sqrt{2}
\end{pmatrix}.\end{eqnarray} 

 The Higgs mutiplets transform according to $\Delta_L \leftrightarrow \Delta_R $ and $ \phi \leftrightarrow \phi^{\dagger} $. The scalar field potential involving the interactions of $\phi$, $\Delta_L$, $\Delta_R$ involves numerous parameters. For reader's convenience, we included the Higgs potenial we have used for our study in Appendix~\ref{A}. Following the notations of Ref.~\cite{Gunion} we define, 
\begin{eqnarray}
\label{para_def}
\rho_{dif}\equiv \rho_3-2(\rho_2+\rho_1),\nn\\ \Delta\alpha\equiv(\alpha_2-\alpha^{'}_2)/2.
\end{eqnarray}
The VEV of the bidoublet $\phi$ and triplets $\Delta_{L,R}$ are given by 
 \begin{eqnarray}
 \braket{\phi} = \dfrac{1}{\sqrt{2}}\begin{pmatrix}
 \kappa_1 & 0 \\ 
 0 & \kappa_2
 \end{pmatrix}, 
 \end{eqnarray}
 \begin{eqnarray}
 \braket{\Delta_{L,R}} = \dfrac{1}{\sqrt{2}} \begin{pmatrix}
 0 & 0 \\ 
 v_{L,R} & 0
 \end{pmatrix}. 
 \end{eqnarray}
 While the bidoublet $\phi$ breaks the SM symmetry $SU(2)_L \times U(1)_Y \rightarrow U(1)_{EM}$, the $SU(2)_R \times U(1)_{B-L} \rightarrow U(1)_Y$ is broken by the triplet VEVs $(v_L,v_R)$ and consequently provides a Majorana mass to the right-handed neutrinos.

\subsection{Vacuum expectation-value scenarios}

 The parameters in the scalar field potential can be severely constrained by various experimental observations. From the $K_L - K_S$ mixing, $W_R$ is constrained to be very heavy ($m_{W_R} > 2.5$ TeV)~\cite{Zhang:2007da,Maiezza:2010ic}. The gauge invariance of $SU(2)_R\times SU(2)_L\times U(1)_{B-L}$ requires the coupling of $W_R$ to both $\Delta_R$ and $\phi$. Consequently $m_{W_R}$ depends on $v_R$, $\kappa_1$ and $\kappa_2$, with both $\kappa_1,\kappa_2$ at the electro-weak (EW) scale. Thus $v_R$ needs to be large ($v_R \gg \kappa_1, \kappa_2$). For the simplicity of calculation we set $\kappa_1 \neq 0$ and $\kappa_2 = 0$.

Thus after fixing $\kappa_2=0$ and $ \kappa_1,v_R \neq 0$, there are two options for the left-handed VEV, $v_L$: (i) $v_L \neq 0$ or (ii) $v_L = 0$. However $v_L$ is highly constrained to be $v_L \ll \kappa$ by the $\rho_{EW}$ parameter, where $\kappa=\sqrt{\kappa^{2}_1+\kappa^{2}_2}$. For $\kappa_2 = 0$, the $\rho_{EW}$ is given in Ref.~\cite{Gunion}, 
\begin{eqnarray}
 \rho_{EW}\equiv\dfrac{M^{2}_{W}}{\cos^{2}\theta_W M^{2}_{Z}} \simeq \dfrac{1 + 2 v^{2}_L/\kappa^{2}_1}{1 + 4 v^{2}_L/\kappa^{2}_1}
\label{eq:rho01}
 \end{eqnarray}
 From the experimentally observed value~\cite{PDG2012}  $\rho_{EW} = 1.0004^{+0.0003}_{-0.0004}$ (at 2$\sigma$), we obtain a bound  $v_L\lesssim 3.5 $  GeV, which is very small compared to $\kappa_1 \approx$ 246 GeV. 
 
 Now let us take a more detailed look at each left-handed VEV scenario.

 \subsubsection{ \texorpdfstring{$v_L \neq 0$}{vLneq0}} 
 \label{vL}

The minimization condition of the Higgs potential in this case sets $\rho_{dif} = 0$. Thus from the mass spectrum of the left Higgs triplet, presented in Appendix \ref{A}, we obtain at the tree level, 
\begin{eqnarray}
m^{2}_{\delta^{0}_L}=m^{2}_{\delta^{0*}_L}=0.
\end{eqnarray}

For $m_{\delta^{0}_L} \ll m_Z$ the $Z$ boson can decay into $\delta^{0}_L \delta^{0*}_L$. Since $\delta^{0}_L$ can only decay to $\nu \nu$ at the tree level, hence the $Z \rightarrow \delta^{0}_L \delta^{0*}_L$ decay is invisible and it becomes constrained by the invisible $Z$ decay width measurement.  
 Using Table~\ref{FR} in Appendix~\ref{C}, we obtain
 \begin{align}
   \Gamma(Z \rightarrow \delta^{0}_L \delta^{0*}_L)=\dfrac{g^2 m_{Z_L}}{48 \pi c^{2}_W}=339 \mev,
   \end{align} 
which violates the experimental uncertainty on the invisible $Z$ width $ \simeq 1.5 \mev$~\cite{PDG2012}. Thus we can conclude that the $v_L \neq 0$ VEV scenario of L-R symmetric models, with further discrete symmetries imposed, is ruled out. The additional discrete symmetries are discussed in Appendix~\ref{A}.
   
    \subsubsection{ \texorpdfstring{$v_L=0$}{vLeq0} }

In this scenario the left-handed Higgs triplet scalars do not mix with their bidoublet counterparts and they are the mass eigenstates by themselves. $\rho_{dif}$ can be arbitrary in this case. The mass spectrum is given as (see Appendix~\ref{A} for details),
    \bea
  m^{2}_{\delta^{0}_L}&=&\dfrac{1}{2}\rho_{dif}v^{2}_R,\nn\\
  m^{2}_{\delta^{+}_L}&=&\dfrac{1}{2}\rho_{dif}v^{2}_R+\dfrac{1}{2}\Delta\alpha\kappa^{2}_1\ =\ m^{2}_{\delta^{0}_L} + \dfrac{1}{2}\Delta\alpha\kappa^{2}_1  , \nn\\
  m^{2}_{\delta^{++}_L}&=&\dfrac{1}{2}\rho_{dif}v^{2}_R+\Delta\alpha\kappa^{2}_1\ =\   m^{2}_{\delta^{0}_L} + \Delta\alpha\kappa^{2}_1.  
  \eea
We shall see in the subsequent sections that these masses, especially their differences, are of much phenomenological interest.   
 
\section{Phenomenology at LHC}
\label{Pheno}

In this section we discuss the VBF production of the doubly charged Higgs boson and their decay channels.

\subsection{Production of the doubly charged Higgs}
\label{3.1}

 \begin{figure}[t]
 \centering
 \includegraphics[width=3.5 in]{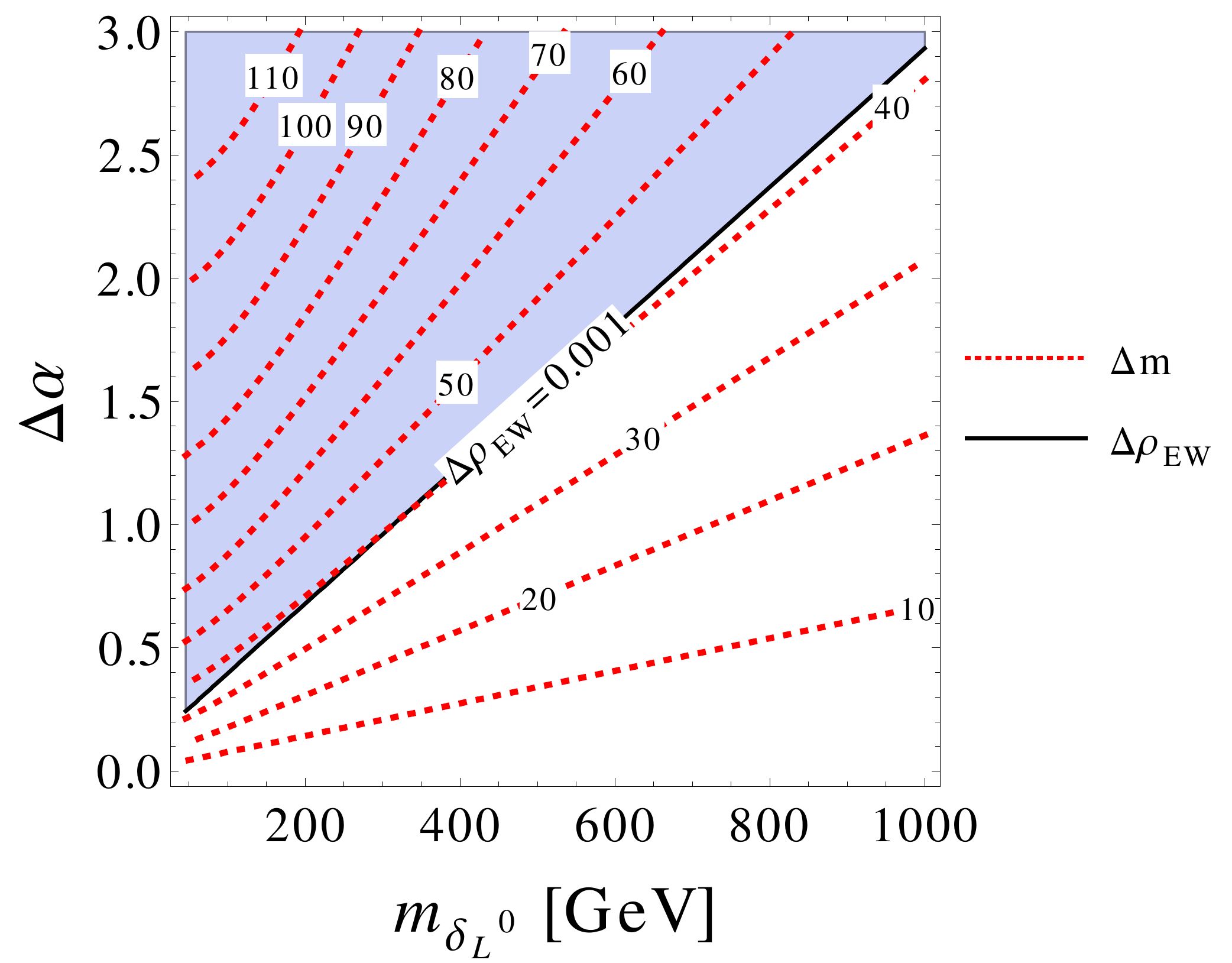} 
\caption{Same value $\Delta m$ contours (dashed) as a function of $m_{\delta^{0}_L}$ and $\delta \alpha$, where $\Delta m = m_{\delta^{++}_L} - m_{\delta^{+}_L} $. The constraint $\Delta \rho_{EW} =0.001 $ is shown by the black thick line. The shaded region is excluded.}
 \label{Param}
\end{figure}
  
The gauge bosons and the left Higgs triplet couplings can be derived from the corresponding Higgs triplet kinetic term of the Lagrangian,
\begin{eqnarray}
\label{Gauge}
\mathcal{L}_{kin}=(\mathcal{D}_{\mu}\Delta_L)^{\dagger}(\mathcal{D}_{\mu}\Delta_L).
\end{eqnarray}
The covariant derivative is defined by,
\begin{eqnarray}
\label{L_Gauge}
\mathcal{D}_{\mu}\Delta_L\equiv i \partial_{\mu}\Delta_L+ \dfrac{g}{2}\dfrac{s_W}{c_W}Y B_{\mu}\Delta_L  +\dfrac{g}{2}[\Vec{\tau}.\Vec{W}_{L_{\mu}},\Delta_L],
\end{eqnarray} 
 where $s_W(c_W)= \sin{\theta_W} (\cos{\theta_W})$, $\theta_W$ being the EW mixing angle. The $\Delta_L$ has a hyper-charge, $Y=2$. 
The Feynman Rules obtained from Eq.~\eqref{Gauge} are listed in Table~\ref{FR} in the Appendix~\ref{C}.

\begin{figure}[t] 
 \includegraphics[width=3 in, height=2 in]{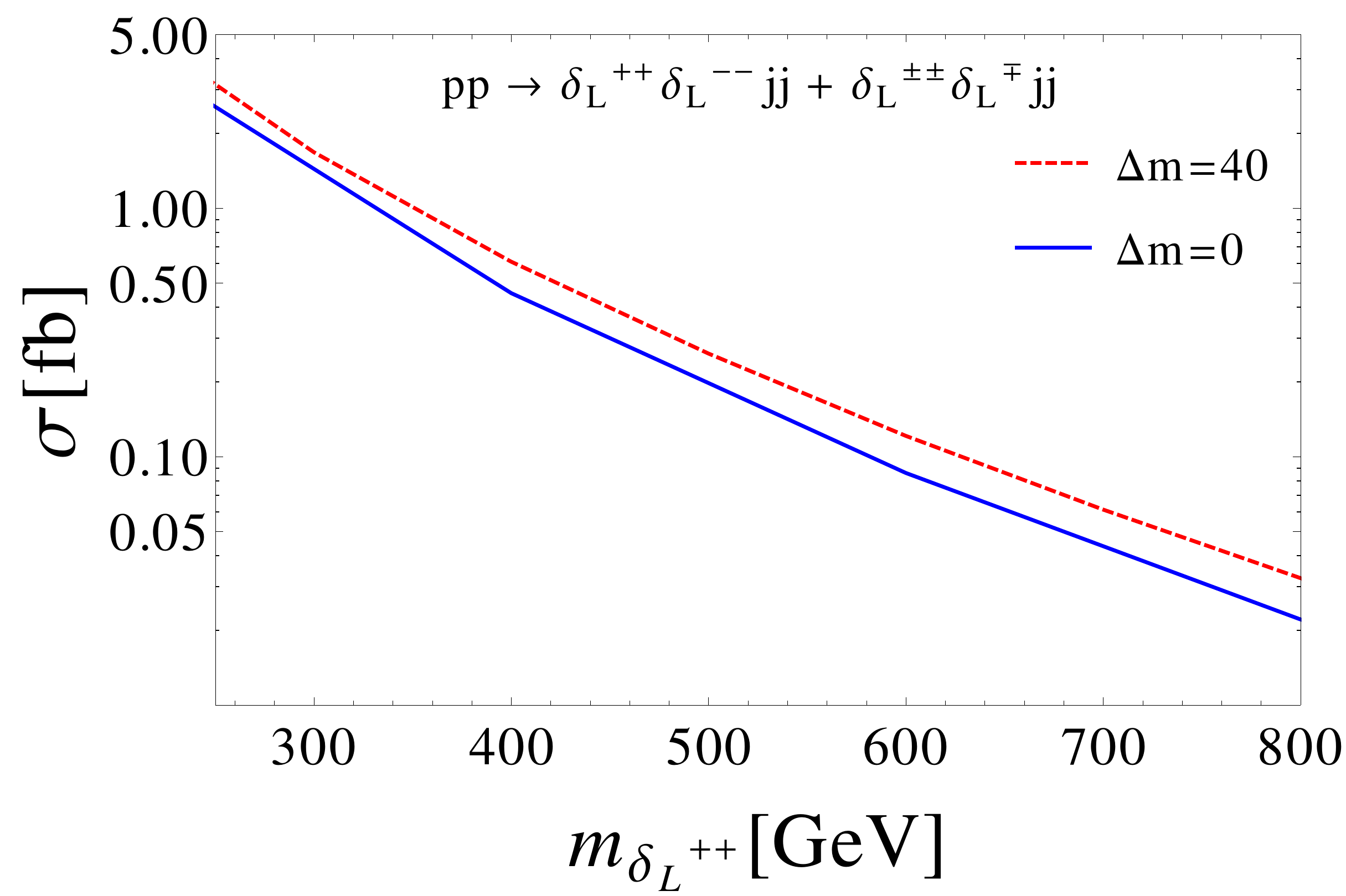}  
 \includegraphics[width=3 in, height=2 in]{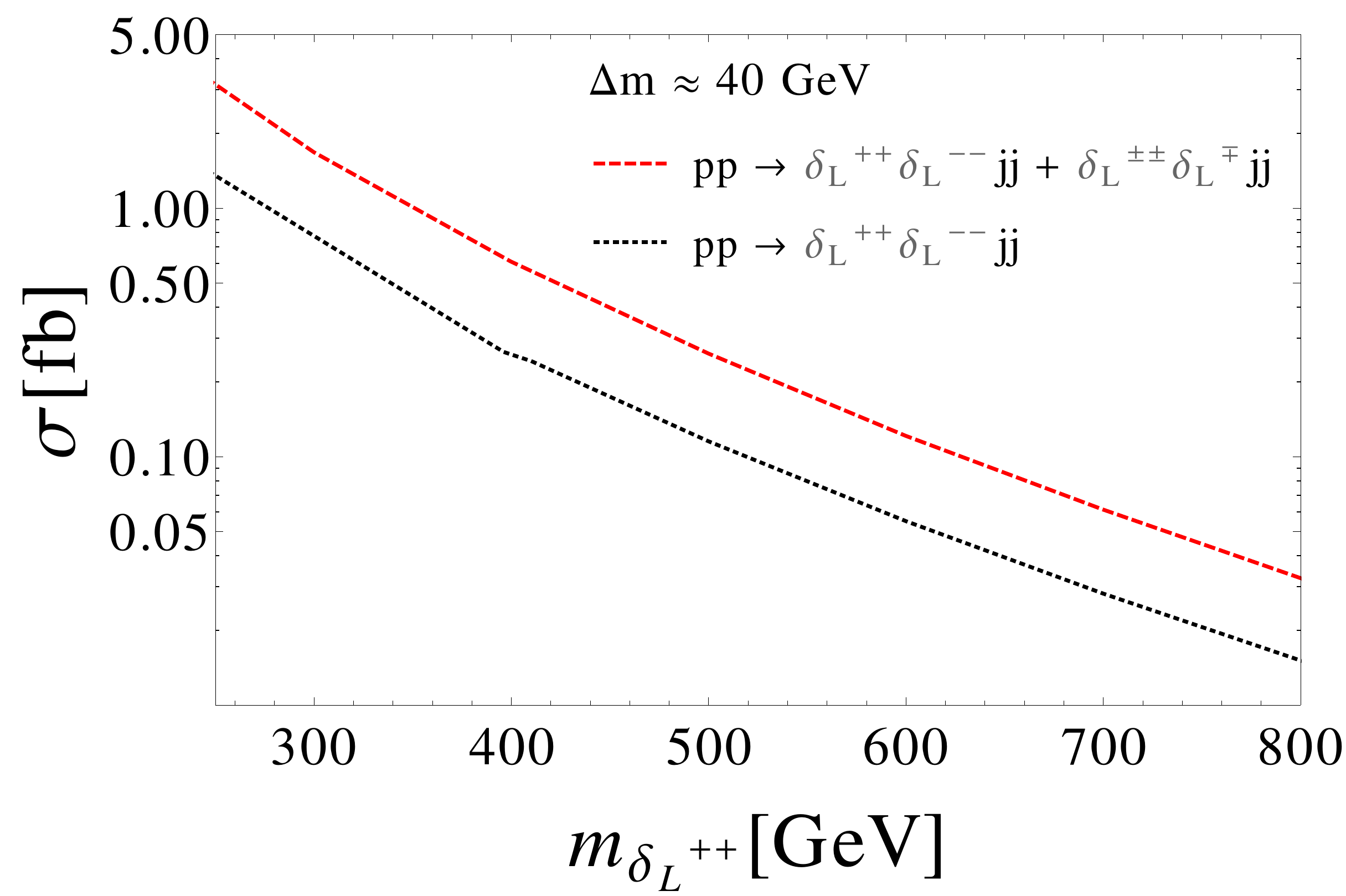} 
  \caption{The cross-section for $pp\rightarrow \delta^{++}_L \delta^{--}_Ljj \ + \ \delta^{\pm \pm}_L \delta^{\mp}_L jj$ at LHC14 for different values of $\Delta m$ [Left panel], where $\Delta m = m_{\delta^{++}_L} - m_{\delta^{+}_L}$, and comparison of the production cross-sections for $pp\rightarrow \delta^{++}_L \delta^{--}_Ljj$ and $pp\rightarrow \delta^{++}_L \delta^{--}_Ljj \ + \  \delta^{\pm \pm}_L \delta^{\mp}_L jj$ at LHC14 [Right panel], keeping $\Delta m \approx 40 \gev$ fixed for all $m_{\delta^{\pm \pm}_L}$.}
\label{Delta m}
\end{figure}

Since $v_L=0$, those gauge boson-scalar couplings proportional to $v_L$ will vanish. Therefore the single production of $\delta^{\pm \pm}_L$  by VBF processes is impossible. Instead we study the production of a pair of doubly charged Higgs $p p \rightarrow \delta^{++}_L \delta^{--}_L j j $. We shall also consider the associated production, $p p \rightarrow \delta^{\pm \pm}_L \delta^{\mp }_L j j $, which increases the production cross-section by a factor of about 2, as shown in Fig.~\ref{Delta m}.
  
{\bf VBF processes:}  
 $ \ $ In a VBF process a pair of Higgs triplet scalars ($\delta^{\pm \pm}_L, \delta^{\pm }_L$) is produced  by the fusion of two virtual vector bosons ($W, Z, \gamma$) radiated by the incoming quarks. Two associated back-to-back `tagging' jets appear in the forward region with a large angular separation $|\Delta \eta_{j_{1}j_{2}}|$~\footnote{Throughout this paper, $j_1 j_2  $ in the subscript denote the dijet pair with the highest invariant mass among all possible pair of jets in the final state. The tagged dijet pairs are sorted among themselves w.r.t their $p_T$.}. A further kinematic cut on the invariant mass of the two jets $M_{j_{1} j_{2}}$, combined with $|\Delta \eta_{j_{1}j_{2}}|$ will be instrumental to reduce SM backgrounds. Since the pair production of two heavy scalars require the incoming partons to be very energetic, the tagged VBF jets should also have high $p_T$. The specific cuts are listed in Section~\ref{Results}.
  
The aforementioned final states can also be produced by DY processes, although the DY production mechanism is not accompanied by the forward-backward jets. Ref.~\cite{delAguila:2013mia} made a comparison between the DY and VBF production and found VBF cross-section is $\sim 5-10 \%$ of the DY cross-section for triplet scalar masses below $1 \tev$. Nevertheless the production of $\delta^{\pm \pm}_L$ by VBF processes and it's subsequent discovery is of utmost importance. From the DY production of $\delta^{\pm \pm}_L$ one can not determine the origin of it. If the interaction of the left Higgs triplet with $W$ boson is absent in the Lagrangian in Eq. \eqref{L_Gauge}, then two photon fusion will be the dominant mechanism for the pair production of $\delta^{\pm \pm}_L$ by VBF processes. Hence the correlation between VBF and DY cross-sections will be different from that if W-couplings were present. Consequently the discovery of $\delta^{\pm \pm}_L$ by means of VBF topological cuts together with their detection in DY channel will shed light on their EW origin.

 With this background let us revisit the mass-spectrum of the left Higgs triplet in the VEV scenario under consideration. The mass splitting $\Delta m = m_{\delta^{++}_L} - m_{\delta^{+}_L} $ can enhance VBF cross-section in the $t$-channel $\delta^{\pm}_L$ exchange, as shown in Fig.~\ref{Delta m}. However this mass difference is also subject to $\rho_{EW}$ constraint via the triplet Higgs' contribution to the $W^{\pm}$ boson mass at loop order, as given in Ref.~\cite{Gunion},
 \begin{eqnarray}
 \Delta \rho_{EW} = \dfrac{2 g^2}{64 \pi^2 m^{2}_W} [f(m_{\delta^{0}_L},m_{\delta^{+}_L}) +f(m_{\delta^{+}_L},m_{\delta^{++}_L})],
 \end{eqnarray}
 where,
 \begin{eqnarray}
 f(x,y)\equiv x^2+y^2-\dfrac{2 x^2 y^2}{x^2 - y^2}\ln (\dfrac{x^2}{y^2}).
 \end{eqnarray}
We recall the fact that the latest experimental value of $\rho_{EW} = 1.0004^{+0.0003}_{-0.0004}$ \cite{PDG2012} allows us to have $\Delta \rho_{EW}  \lesssim 0.001$ at $2\sigma$. This yields a severe bound of $\Delta m_{max} \sim 40 \gev$. The bound on the mass splitting as a function of  $m_{\delta^{0}_L}$ and $\Delta \alpha$ has been presented in Fig.~\ref{Param}. This bound also applies for the mass splitting between $\delta^{+}_L$ and $\delta^{0}_L$. It is worthwhile to mention here that a bound on $\Delta \alpha$ may arise from the mass bounds on the FCNC scalars \cite{Gunion}. For a fixed $\Delta \alpha$ and $\Delta m$ we can set an upper bound on the mass of $\delta^{0}_L$. The dependence of the cross-section on $\Delta m$ is illustrated in Fig.~\ref{Delta m}. Here we have plotted the variation of the cross-section as a function of $m_{\delta^{++}_L}$ for $\Delta m = 0$ and $40$ cases. 
 
\subsection{Decay of the doubly charged Higgs}

Here we will show that with $v_L =0 $, $\delta^{++}_L$ dominantly decays to a pair of same sign leptons ($e, \mu$, $\tau$) that can give rise to an unique collider signature.
 Since $B-L$ charge is 2 for the triplet, their Yukawa couplings are given by the Lagrangian,
\begin{eqnarray}
\label{Yukawa}
\mathcal{L}_Y=i h^{M}_{ij}\psi^{T}_{iL}\mathcal{C}\tau_2\Delta_L\psi_{jL} + c.c.,
\end{eqnarray}
where $i,j$ are the lepton generation indices, and $\mathcal{C}$ is the charge-conjugation operator.  $\psi_L$ is the left-handed lepton doublet. Various constraints may arise on the Yukawa couplings from different experiments. These constraints on $h^{M}_{ij}$ have been discussed at a great detail in Refs.~\cite{Gunion,Huitu,Quintero:2012jy}. Here we only list the relevant ones for our study.

The Bhaba scattering yields a constraint~\cite{Huitu} on $h^{M}_{ee}$, 
\begin{eqnarray}
(h^{M}_{ee})^2\lesssim 9.7\times 10^{-6} \left(\frac{m_{\delta^{++}_L}}{\text{GeV}}\right)^2.
\end{eqnarray}   

The moun $g$-2 receives contribution from the $\delta^{++}_L$ ~\cite{Gunion, Queiroz:2014zfa}. Using the current measurement, $\Delta a_{\mu}=a^{exp}_{\mu}-a^{SM}_{\mu}=288\times10^{-11}$~\cite{PDG2012}, we obtain,
\begin{eqnarray}
(h^{M}_{\mu \mu})^2\lesssim 1.5\times 10^{-5} \left(\frac{m_{\delta^{++}_L}}{\text{GeV}}\right)^2.
\end{eqnarray}

From the upper limit of BR$(\mu \rightarrow e \gamma)$~\cite{PDG2012,Huitu} we get,
\begin{eqnarray}
 h^{M}_{ee}\cdot h^{M}_{\mu \mu} \lesssim 5.8\times10^{-5} \left(\frac{m_{\delta^{++}_L}}{\text{GeV}}\right)^2.
 \end{eqnarray} 

The non-diagonal coupling $h^{M}_{\mu e}$ can break the lepton number of each generation and is the most stringently constrained~ \cite{PDG2012,Gunion}. Between the first two generation, we find,
\begin{eqnarray}
\label{hemu}
h^{M}_{\mu e}\cdot h^{M}_{ee} \lesssim 2\times10^{-11} \left(\frac{m_{\delta^{++}_L}}{\text{GeV}}\right)^2.
\end{eqnarray}

Finally, no constraint on $h^{M}_{\tau \tau}$ is available yet. 
 
$\Delta L=2$ decay rates into different flavor combinations depend on the relative strength of the Yukawa couplings $h^{M}_{ij}$. In this paper we consider two sample cases: (i) BR($\delta^{++}_L \rightarrow e^+ e^+$) = 50\%, BR($\delta^{++}_L \rightarrow \mu^+ \mu^+$) = 50\% and (ii) BR($\delta^{++}_L \rightarrow \tau^+ \tau^+$) = 100\%. For Case (i), we assume a diagonal $h^{M}_{ij}$.

 As $v_L =0 $, the decay processes $\delta^{++}_L \rightarrow W^{+} W^{+}$ and $\delta^{++}_L \rightarrow \delta^{+}_L \delta^{+}_L$ are forbidden. The constraints on the mass-splittings rule that the decays $\delta^{++}_L \rightarrow \delta^{+}_L W^{+}$, $\delta^{++}_L \rightarrow \delta^{+}_L \delta^{+}_L \delta^{0}_L$ and $\delta^{++}_L \rightarrow W^{+} W^{+} \delta^{0}_L$ can only occur virtually. These virtual decays have a subdominant branching ratio compared to that of the dilepton decay channel, hence we ignore them in this paper.

\section{Results}
\label{Results}
    
Here we present the significance of a potential LHC signal from the two $\delta^{\pm \pm}_L$ decay scenarios, $\delta^{\pm \pm}_L\rightarrow l^{\pm}l^{\pm}$ ($ll=50\%~ee,~50\%~\mu\mu$) and $\delta^{\pm \pm}_L\rightarrow \tau^{\pm} \tau^{\pm} (100\%)$. 
The L-R symmetric model is implemented with the {\tt{FeynRules v1.7.200}}~\cite{FeynRules} package. The signal and background events are generated using {\tt{MADGRAPH5}}~\cite{Alwall:2011uj} followed by showering and hadronization  by {\tt{PYTHIA}}~\cite{Sjostrand:2006za} and the detector simulation by {\tt{PGS4}}~\cite{pgs}.

\subsection{Like-sign light lepton pairs}
 \label{subsec:lightlepton} 
  \begin{figure}[!ht]
 \centering
  \includegraphics[width=3.5in]{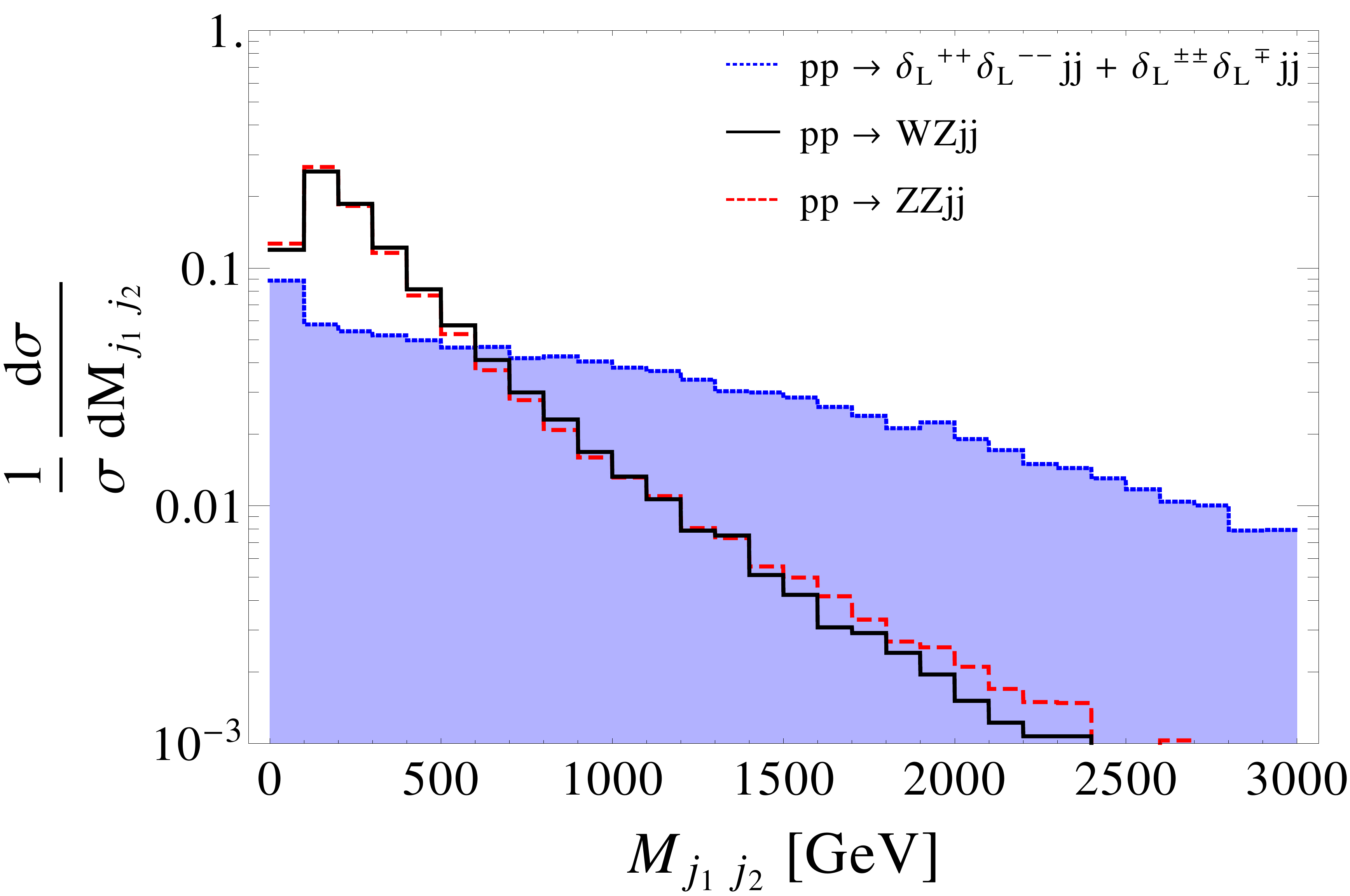}
  \caption{Distribution of the dijet invariant mass $M_{j_{1} j_{2}}$ normalized to unity for the tagging jet pair $(j_1,j_2)$ for the signal and all the backgrounds before the VBF cuts are applied. The signal is inclusive of the processes $\delta^{++}_L \delta^{--}_L jj$ and $\delta^{\pm \pm}_L \delta^{\mp}_L jj$ for the benchmark point ($m_{\delta^{++}_L},m_{\delta^{+}_L}$) = ($460 \gev,420 \gev$) at $\sqrt{s}=14 \tev$. The tagged jets $(j_1,j_2)$  being the pair of jets with largest dijet invariant mass among all final state jet pairs. $j_1$ and $j_2$ have been sorted among themselves by their $p_T$.}
\label{emuMjj}
\end{figure}
 
We first consider the decay of $\delta^{\pm \pm}_L$ into like-sign light lepton pairs ($e , \mu$). The production processes $p p \rightarrow \delta^{\pm \pm}_L \delta^{\mp}_L j j$ and $p p \rightarrow \delta^{++}_L \delta^{--}_L j j$ lead to a final state of at least three leptons ($\geq 3 l$) and two or more jets ($\geq 2 j$).

The CMS analysis~\cite{Chatrchyan:2012ya} has set the latest bound on $m_{\delta^{\pm \pm}_L}$ from the 3 lepton channel to be  $444 \gev$ for a 100\% decay to $ee$ and $459 \gev$ for 100\% decay into $\mu \mu$. The corresponding bounds for a 4 lepton final states are respectively $382 \gev$ for $ee$ and $395 \gev$ for $\mu \mu$. In comparison the ATLAS bound~\cite{Aad:2012xsa} for the $\geq 3 e/ \mu$ channel is $330 \gev$, and their bounds from the 4 lepton channel~\cite{Aad:2012cg,Quintero:2012jy} are $409 \gev$ and $398 \gev$ for $ee$ and $\mu \mu$, respectively.
  
The major SM backgrounds for our final state are $pp \rightarrow W Z j j$ and $pp \rightarrow Z Z j j$. The kinematic cuts imposed to reduce these background are as follows:  

(1) Basic Cuts: The signal and background events are preselected with the requirement of at least 2 Jets with $p_{T j} > 30 \gev$ and $|\eta_j| < 5$. The subsequent cuts applied on the pre-selected events are optimized to maximize the signal significance, $S/\sqrt{S+B}$, where $S$ and $B$ denote signal and background rates. 

(2) VBF Cuts: Denoting $j_1,j_2$ to be the pair of jets with the highest $M_{jj}$, we require $(i)\ p_{T}(j_1), p_{T}(j_2) > 50 \gev$, $(ii)\ \eta_{j_1}*\eta_{j_2}<0$, $(iii)\ |\Delta \eta_{j_1,j_2}| > 4 $ and $(iv)\ M_{j_1 j_2} > 500 \gev$. We find from Fig.~\ref{emuMjj} that the $M_{j_1j_2}$ of the backgrounds fall below the same of the signal above $500 \gev$~\footnote{For all the figures presented in this paper, the signal and background kinematic distributions are not stacked. }. 

(3) $\geq 3$ leptons : At least 3 light leptons ($l=e,\mu$) with $p_{T}(l) > 10 \gev$ and $|\eta_{l}| < 2.5$. This cut reduces the background by a factor of $10^{-2}$ due to the low $W \rightarrow l \nu$ and $Z \rightarrow l l$ branching fractions.

\begin{figure}[!ht]  
\centering 
 \includegraphics[width=3 in, height=2 in]{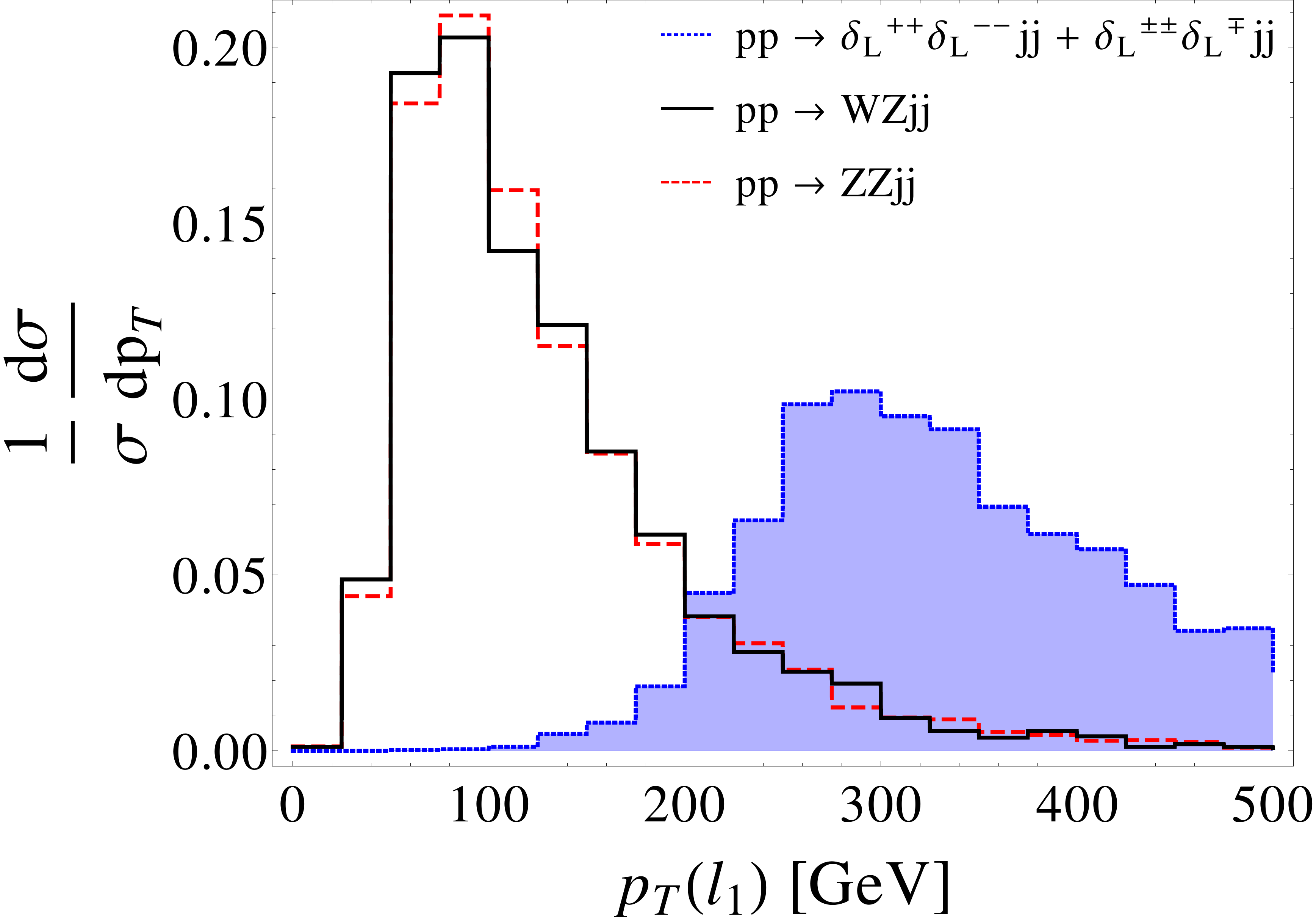}
 \includegraphics[width=3 in, height=2 in]{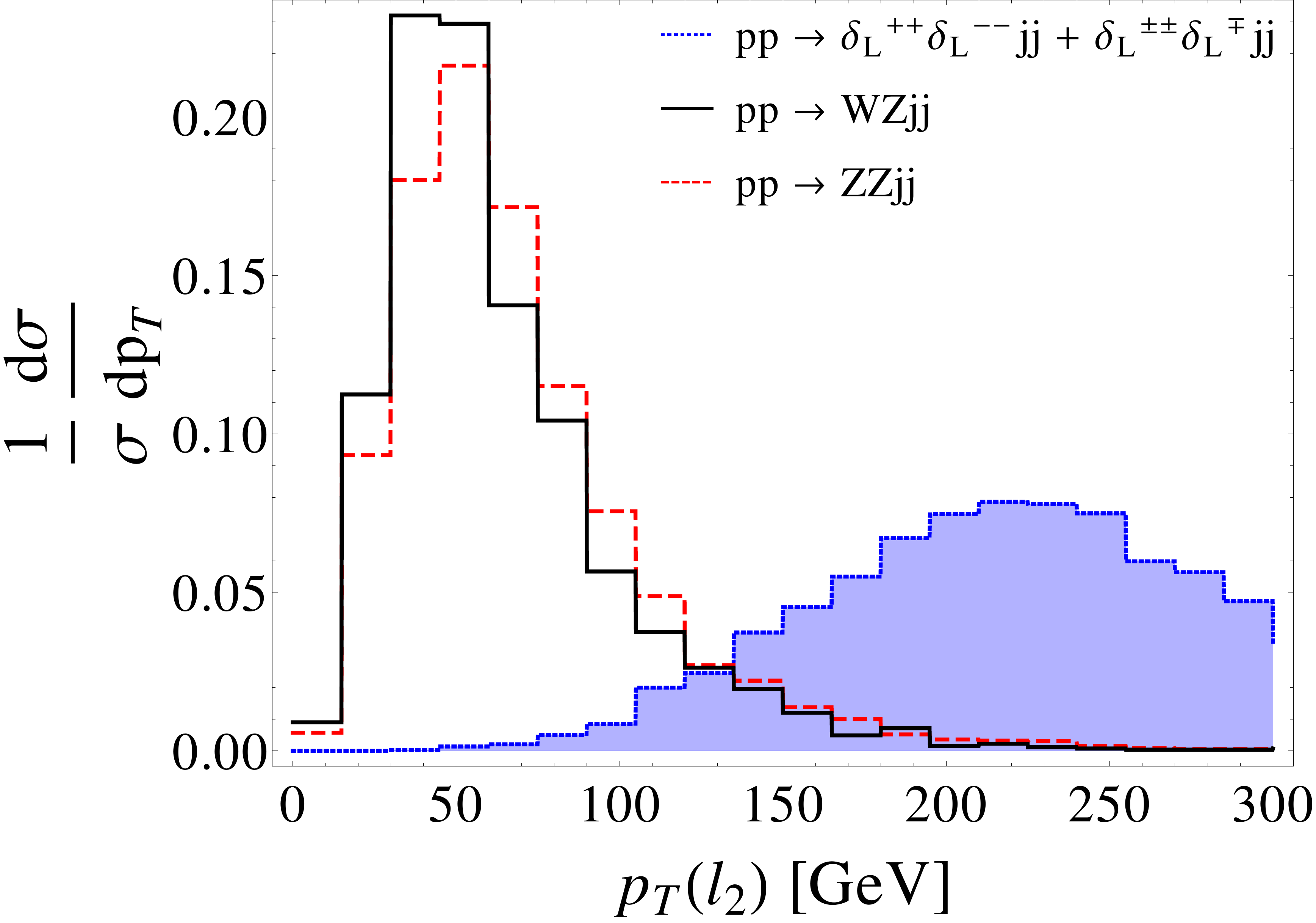}\\ 
 \includegraphics[width=3 in, height=2 in]{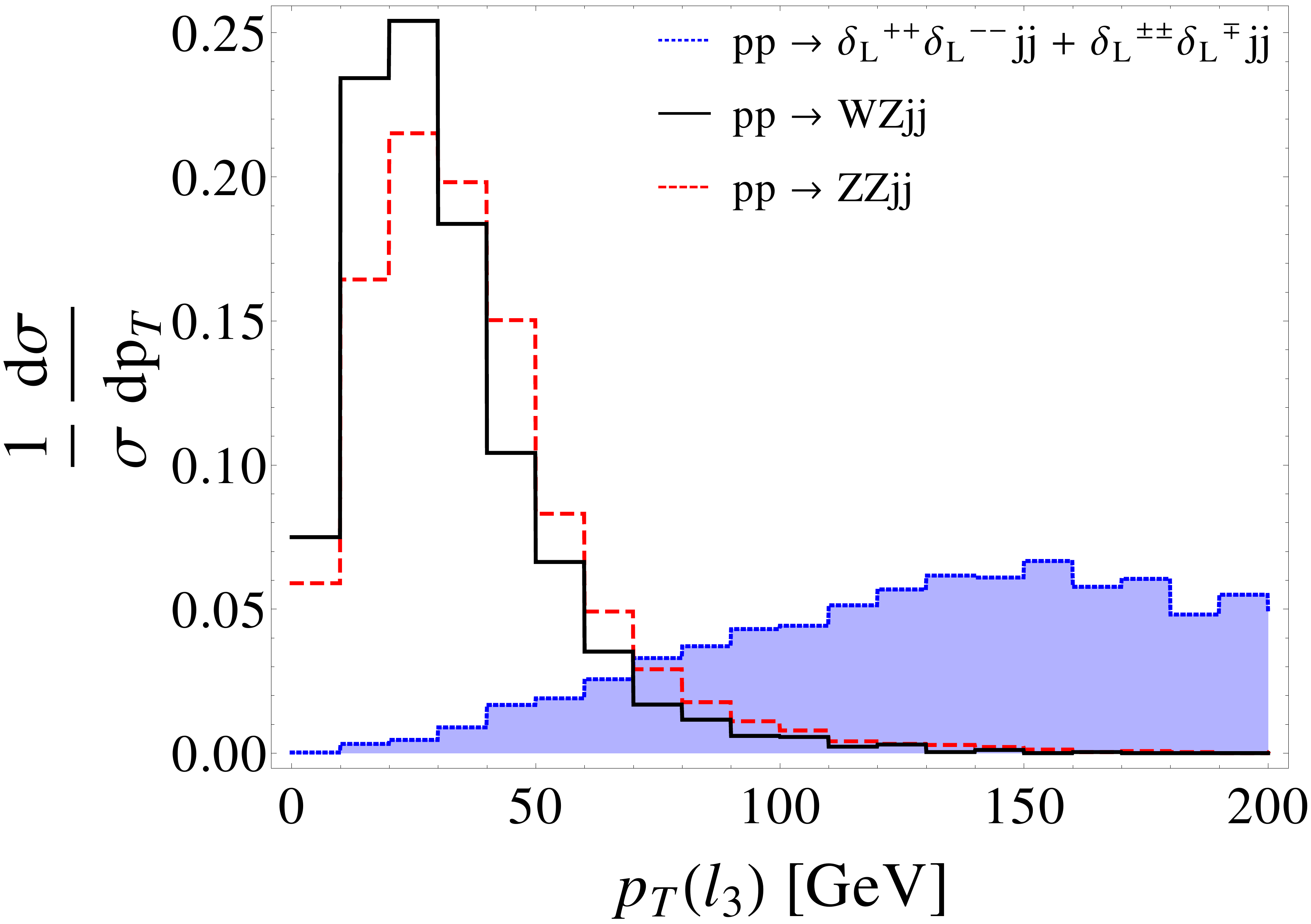} 
 \includegraphics[width=3 in, height=2 in]{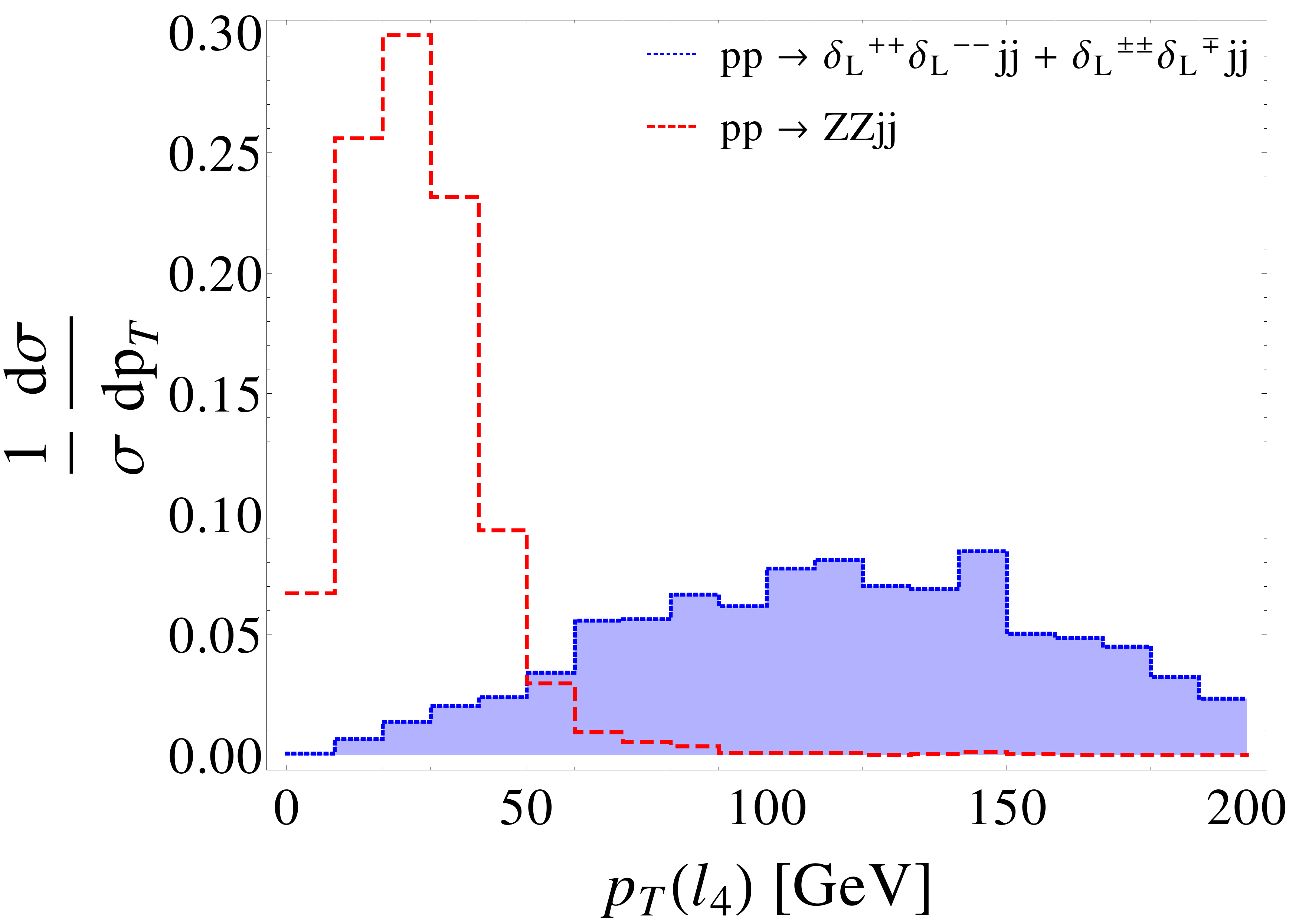} 
  \caption{$p_T$ distribution of the leptons, normalized to unity, in the $\geq 3 l$ + $\geq 2 j$ final state for the signal [($m_{\delta^{++}_L},m_{\delta^{+}_L}$) = ($460 \gev,420 \gev$)] and SM backgrounds at $\sqrt{s}=14 \tev$.}
\label{emuPT}
\end{figure}

(4) Lepton $p_T$ cuts: The $p_T$ distribution of the leptons of the signal and the backgrounds are shown in Fig.~\ref{emuPT}. The figure clearly shows that the leptons arising from $\delta^{\pm \pm}_L$ decay are more energetic than the leptons coming from $W,Z$ bosons. This allows us to impose stringent $p_T$ cuts on the leptons as follows: $p_T(l_1) > 120 \gev$, $p_T(l_2) > 100 \gev$, $p_T(l_3) > 50 \gev$, and $p_T(l_4) > 30 \gev$. These cuts are instrumental to reduce  backgrounds by an order of magnitude. 

 (5) $Z$ veto: Opposite-sign dilepton invariant mass, $|M_{l^+l^-} - m_Z| > 10 \gev$  removes the $ZZjj$ background effectively.
 
 (6) $\delta^{++}_L$ mass window cut: Like-sign dilepton invariant mass, $|M_{l^{\pm}l^{\pm}} - m_{\delta^{++}_L}| < 10\% ~m_{\delta^{++}_L}$.

\begin{table}[!htp] 
\begin{tabular}{|c |c |c |c |c |} 

\hline  
     $(m_{\delta^{++}_L}, m_{\delta^{+}_L}) $  & Selection  & Signal  & ZZjj  & WZjj 
     \\
$[\gev]$  & Cuts  & [fb] & [fb] & [fb]   
\\          
          \hline    \hline  
                     
 &  Basic cuts  & $0.1540 \pm 0.0011$ & $585.9 \pm 1.4$  & $3513 \pm 8$ 
  \\
 &  VBF & $0.0403 \pm 0.0005 $ &  $39.98 \pm 0.36$ & $211.8 \pm 2.1$ 
  \\
 $(460,420)$       &  $\geq$ 3 leptons     & $0.0317 \pm 0.0005$   & $0.2131 \pm 0.0028$  & $1.702 \pm 0.033$ 
  \\
 &  lepton $p_T$ cuts  & $0.0301 \pm 0.0005$  &  $0.0126 \pm 0.0007$  & $0.1015 \pm 0.0080$
  \\                     
 &  $Z$-veto & $0.0291 \pm 0.0005$    & $0.0005 \pm 0.0001 $   &  $0.0057 \pm 0.0019$  
\\   
 &  $\delta^{++}_L$ mass window & $0.0285 \pm 0.0005$ & $0.0001 \pm 0.0001$ & $0.0002 \pm 0.0002$  
 \\  \hline 
\end{tabular} 
\caption{Summary of the signal and the background cross-sections and corresponding statistical errors at our chosen benchmark point, after each kinematical cut in the light lepton decay scenario. The LHC energy is 14 TeV.} 
\label{emuX-sec}
\end{table}

\begin{figure}[!ht]
 \centering
  \includegraphics[width=2.75 in]{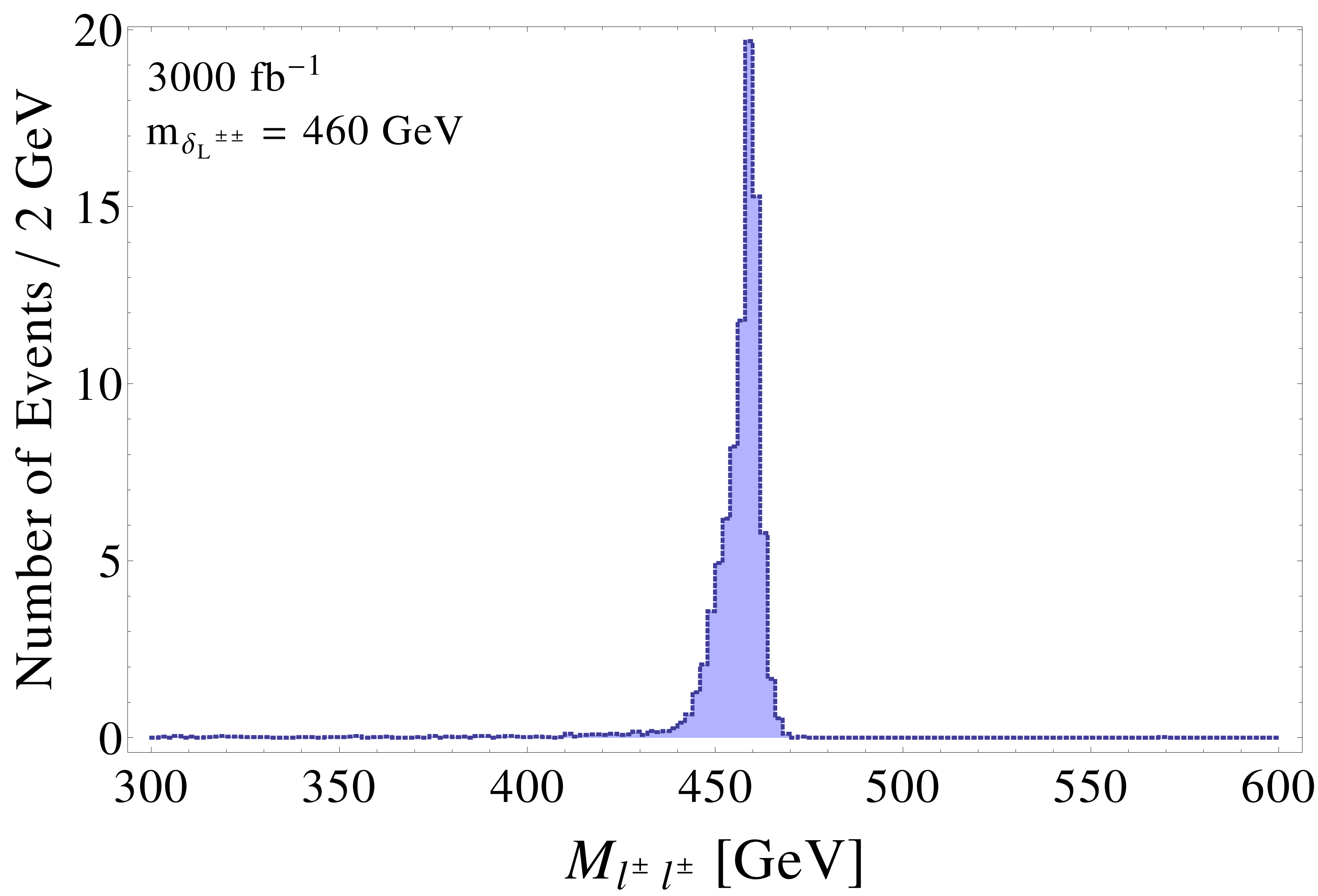} 
  \includegraphics[width=3 in]{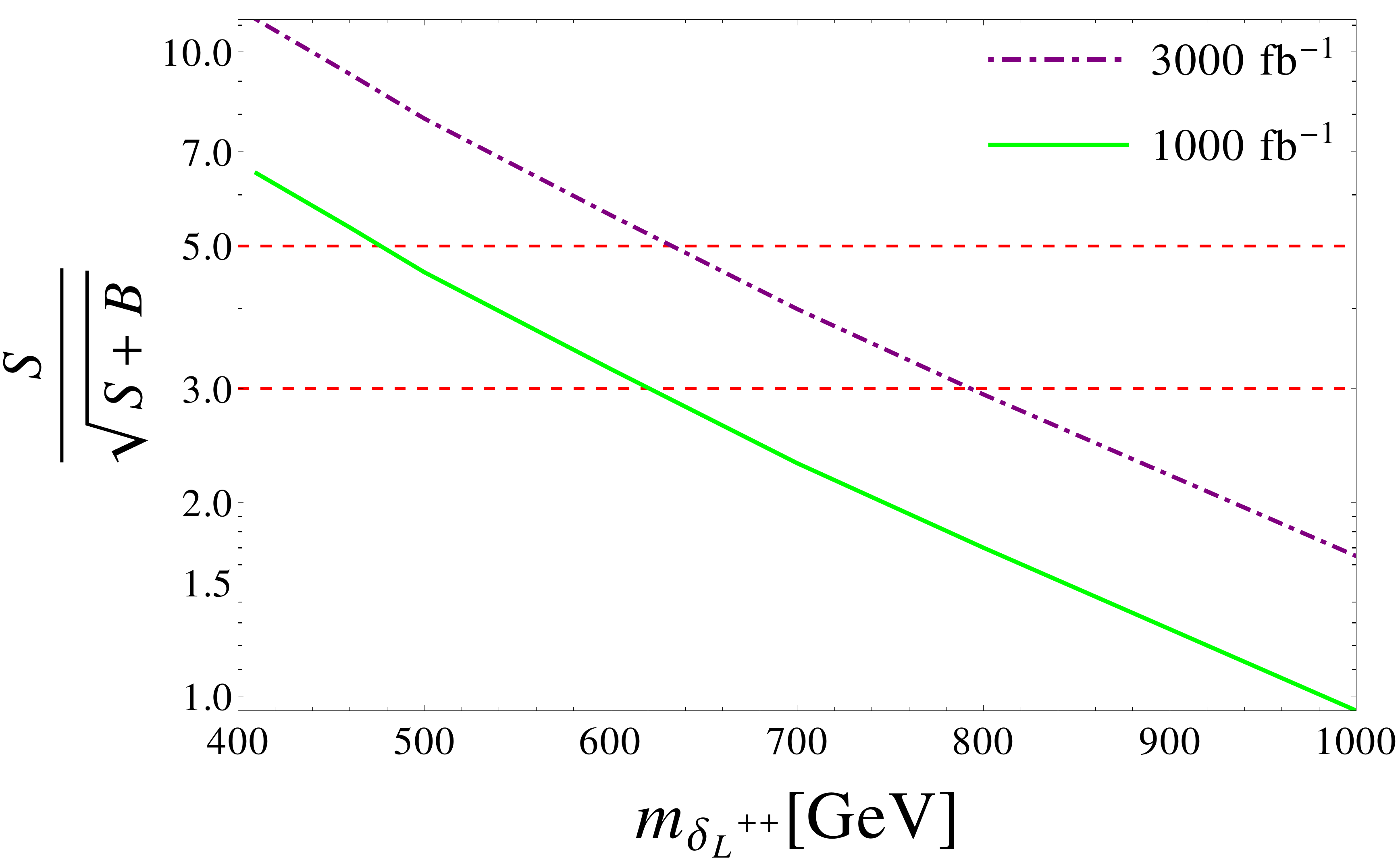} 
  \caption{The signal invariant mass distribution [Left panel] of same sign dilepton pairs in the $\geq 3 l$ + $\geq 2 j$ final state, after applying all the kinematic cuts. Since the number of background events drop to less than $1$ after all the cuts, they have not been included in this figure. The significance [Right panel] curves assume $1000$ fb$^{-1}$ (solid) and  $3000$ fb$^{-1}$ (dot-dashed) LHC luminosities at 14 TeV. The 3 and 5 $\sigma$ level are shown as horizontal lines and $\Delta m$ is kept to be $\approx 40 \gev$.}
\label{emuMll}
\end{figure}

The reconstructed invariant mass of the  like-sign dilepton pairs after all the cuts is shown in Fig.~\ref{emuMll} for $m_{\delta^{++}_L} = 460 \gev$. One can readily observe the distribution peaks sharply at $m_{\delta^{\pm \pm}_L}$. Table~\ref{emuX-sec} gives the signal and background cross-sections after applying each cut. The significances ($S/\sqrt{S+B}$) have been plotted in Fig.~\ref{emuMll} as a function of $m_{\delta^{++}_L}$, keeping $\Delta m$ fixed at $\sim 40 \gev$, for upcoming LHC luminosities at $1000$ fb$^{-1}$ and $3000$ fb$^{-1}$. We find that at $3 \sigma (5 \sigma)$ level the $m_{\delta^{++}_L} $ can be probed up to $ 620 (480) \gev$ for 1000 fb$^{-1}$ luminosity, and $800 (640) \gev$ with 3000 fb$^{-1}$.
 
\begin{figure}[!ht]  
\centering 
 \includegraphics[width=3 in, height=2 in]{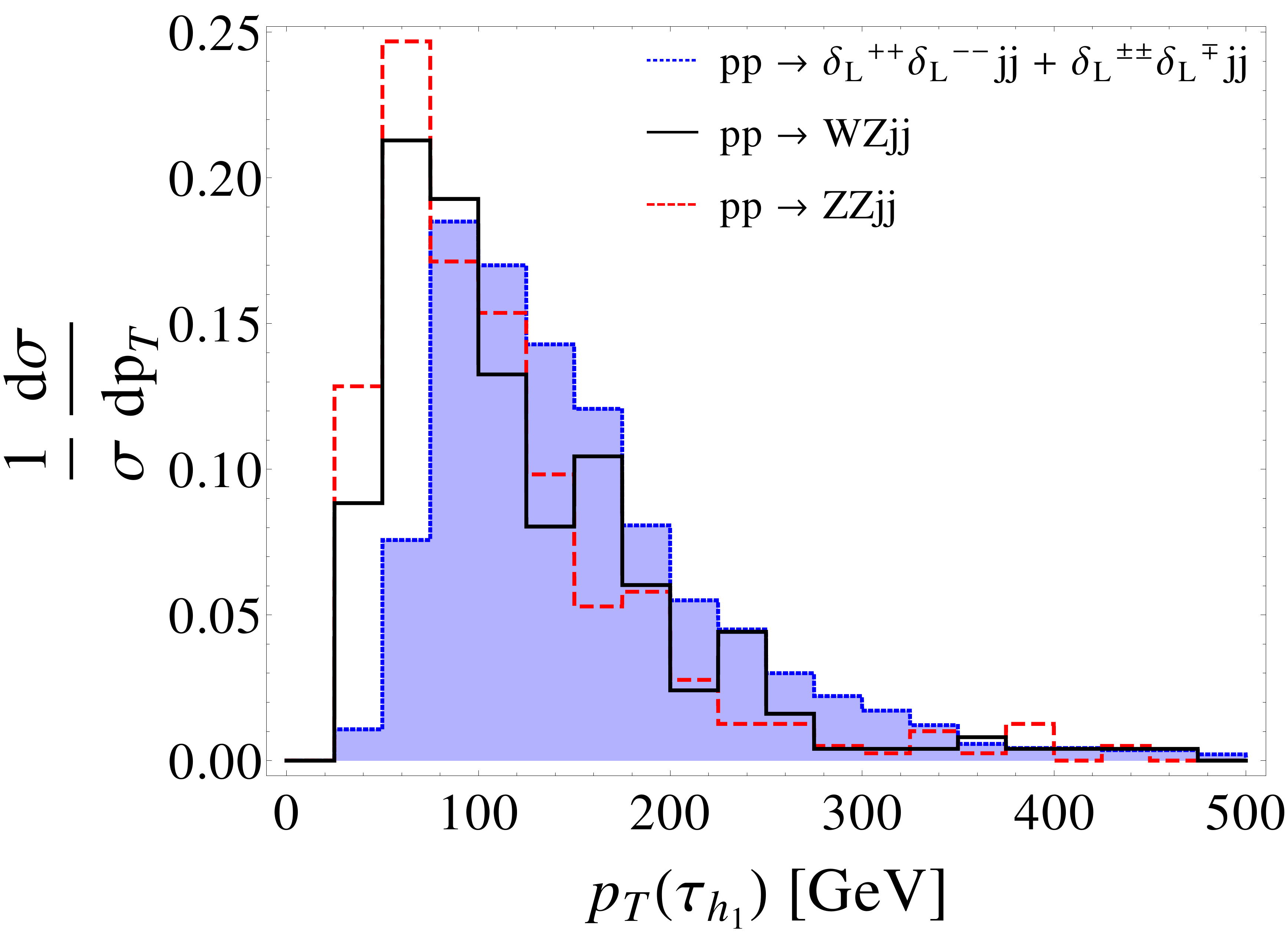}
 \includegraphics[width=3 in, height=2 in]{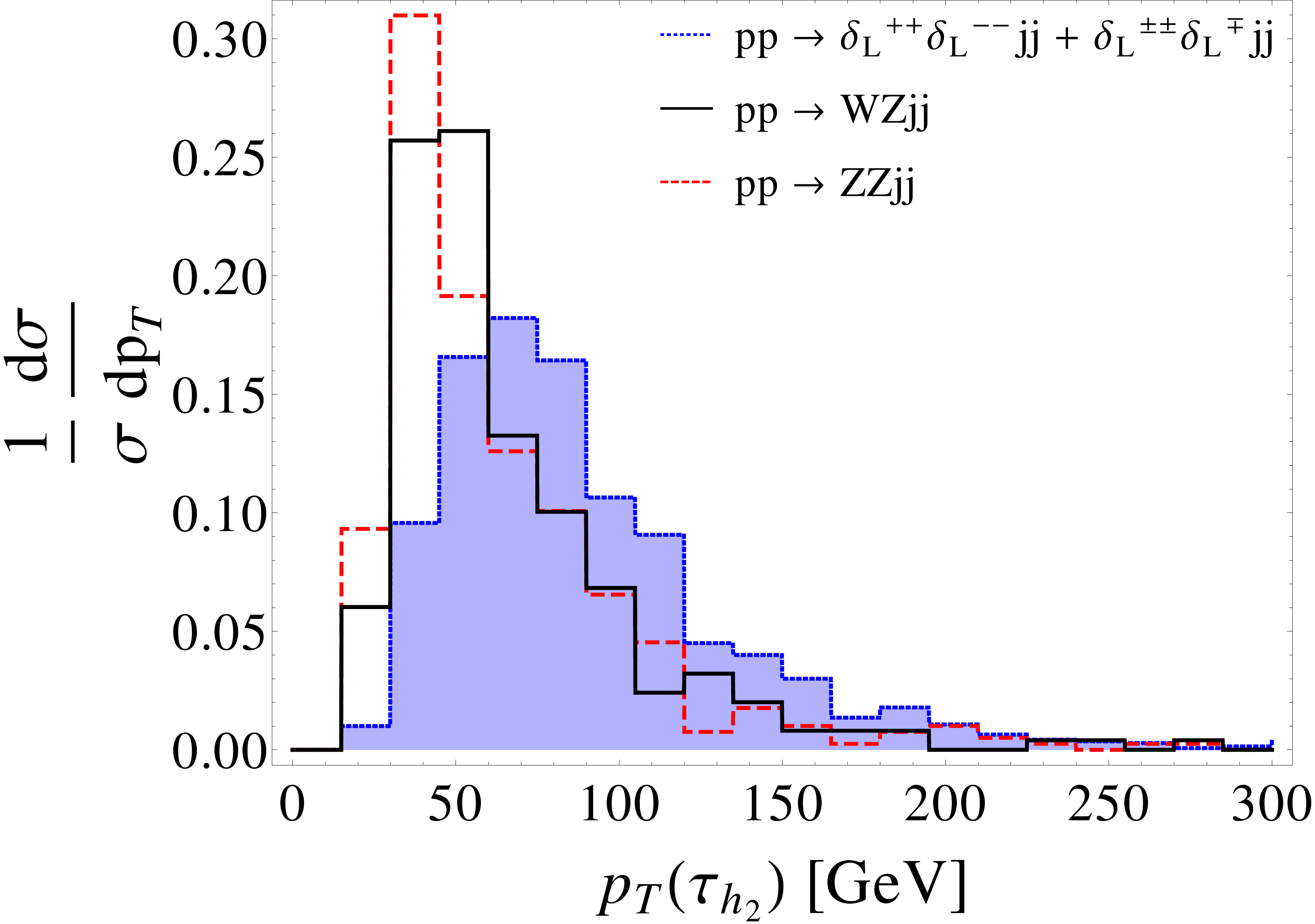}
 \includegraphics[width=3 in, height=2 in]{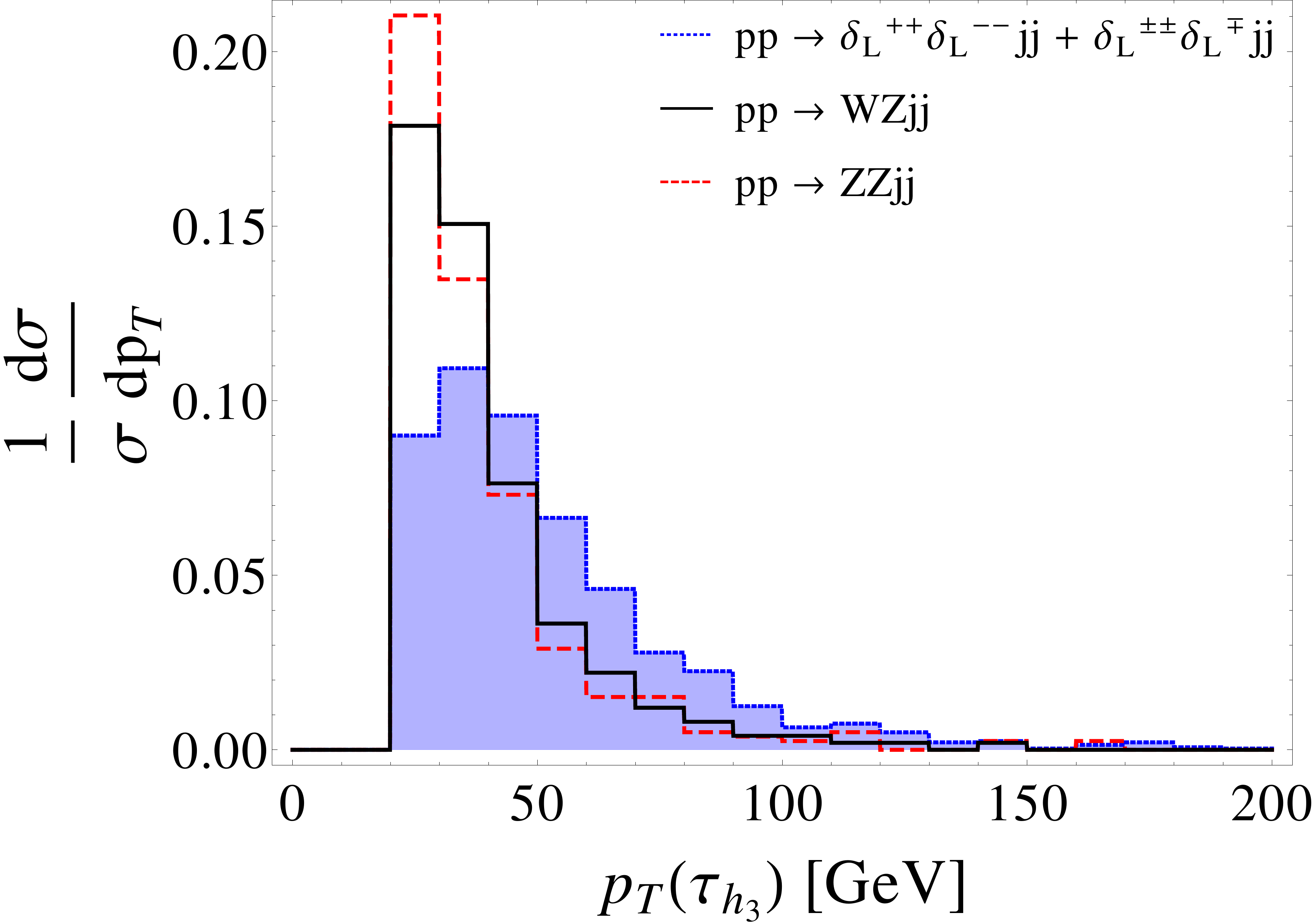}   
  \caption{$p_T$ distribution of the $\tau_h$s, normalized to unity, in the $\geq 3 \tau_h$ + $\geq 2 j$ + $\met$ final state for the signal [($m_{\delta^{++}_L},m_{\delta^{+}_L}$) = ($210 \gev,170 \gev$)] as well as SM backgrounds at $\sqrt{s}=14 \tev$. For all the plots the $\tau$ ID efficiency and $jet \rightarrow \tau_h$ fake rate are considered to be $70 \%$ and $2 \%$ respectively.}
\label{tauPT}
\end{figure}

  \subsection{Like-sign tau pairs}
  
In this subsection we shall present the VBF search strategy in the multi-$\tau$ final state. The search with $\tau$ leptons is the most difficult amongst the three generation of leptons because of low  $\tau$ identification efficiency at the LHC. The $\tau$ lepton tagging requires the $\tau$ to decay hadronically, denoted  by `$\tau_h$'. Similar to the light lepton case, the signal final state contains $\geq 3 \tau_{h}$ + $\geq 2 j$ together with $\met$. Existing CMS bounds~\cite{Chatrchyan:2012ya} for the $\tau \tau$ final state (for the DY process) are $204 \gev$ and $169 \gev$ in the 3 and 4 lepton channel, respectively. The ATLAS does not offer a limit for this scenario. 

Tagging the $\tau_h$ involves the $\tau_h$ identification (ID) $\epsilon$ and the fake rate $f$ coming from jets. We look at the case $\epsilon = 50 \%$, $f= 1 \%$, both $\epsilon$ and $f$ flat over $p_T > 20 \gev$~\cite{CMS:2011msa}, and an alternative enhanced efficiency at $\epsilon = 70 \%$, which could be achievable in future LHC searches \cite{CMS:70_2}, with a higher fake rate $f= 2 \%$~\footnote{{\tt{PGS4}}, by default, has a $\tau_h$ identification (ID) efficiency ($\epsilon$) of $~30-40\%$. To study the impact on this search due to $\tau_h$ ID performance, an object reconstructed by visible particles from the hadronic decay of $\tau$ in {\tt{PYTHIA}}, is used at a rate of ID efficiency ($50 \%$ or $70 \%$). Correspondingly each {\tt{PGS4}} jet object is misidentified as a $\tau_h$ object at a probability of $1 \%$ or $2 \%$ (fake rate $f$ in the text).}. The kinematical cuts imposed are listed as follows,
  
(1) Basic and (2) VBF cuts: Same as in the light leptons case. See in Section~\ref{subsec:lightlepton}. Rest of the cuts are similarly optimized to maximize the signal significance, $S/\sqrt{S+B}$.

(3) $\geq 3$ tagged $\tau_h$ : We have selected events with at least 3 $\tau_h$s in the final state with $p_T(\tau_h) >20 \gev$ and $\eta(\tau_h)<2.3$, which are required~\cite{CMS:2011msa} for our assumed $\epsilon$ and $f$.

(4) $\tau_h \ p_T$  cuts: since a $\tau$ loses part of its $p_T$ as $\met$, illustrated in Fig.~\ref{tauPT}, we use a softer $p_T$ cuts: $p_T(\tau_{h_ 1}) > 50 \gev$, $p_T(\tau_{h_2}) > 50 \gev$ and $p_T(\tau_{h_3}) > 30 \gev$ and no $p_T$ cut for additional $\tau_{h}$s.  This cut achieves a $ 75 \%$ cut efficiency for the signal, and reduces the background by a factor of $2$.

(5) $\met$ cut: $\met > 50 \gev$. 
  
\begin{figure}[!ht]  
\centering 
 \includegraphics[width=3. in]{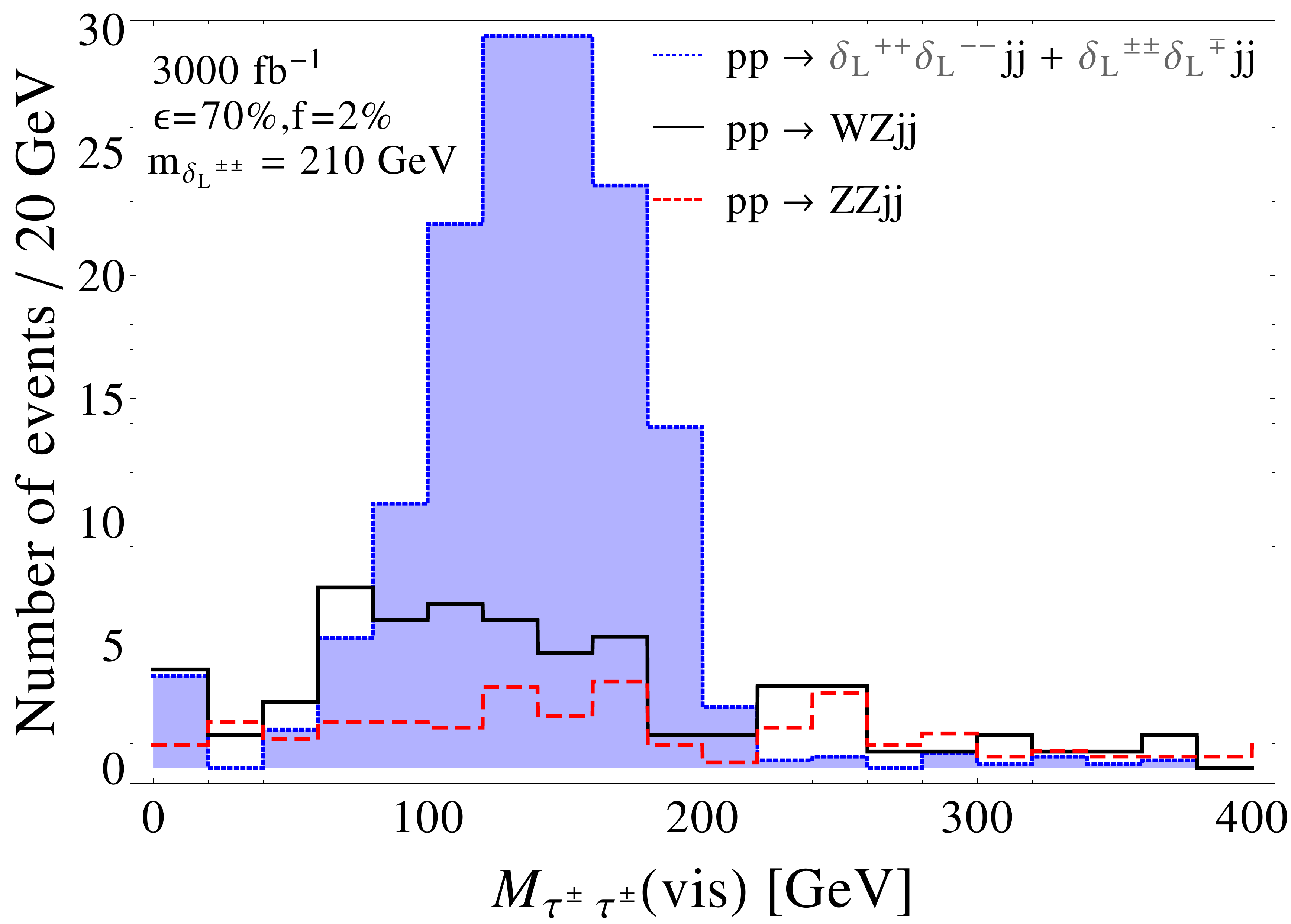} 
\caption{The visible mass of the like-sign $\tau_h$  pair for the signal (dotted) and backgrounds (solid and dashed) for the $\geq 3 \tau_h$ + $\geq 2 j$ + $\met$ final state, after all the kinematic cuts.  ($m_{\delta^{++}_L},m_{\delta^{+}_L}$) = ($210 \gev,170 \gev$) with $\epsilon=70\%$ and $f=2\%$ at $14 \tev$.}
\label{MT}
\end{figure}

Because the $\tau$ decays into the invisible $\nu_\tau$, the invariant mass of the same-sign $\tau$ pair cannot be reconstructed completely.  
Fig.~\ref{MT} shows the visible mass, $M_{\tau^{\pm} \tau^{\pm}}(vis)$, of the like-sign $\tau_h$ pair from $m_{\delta^{++}}=210$ GeV. Due to the loss of visible $p_T$, the $\tau \tau$ mass show a broad distribution that centres at 140 GeV, below the parent $\delta^{++}_L$ mass.
 
\begin{figure}[!ht]  
\centering 
 \includegraphics[width=3 in]{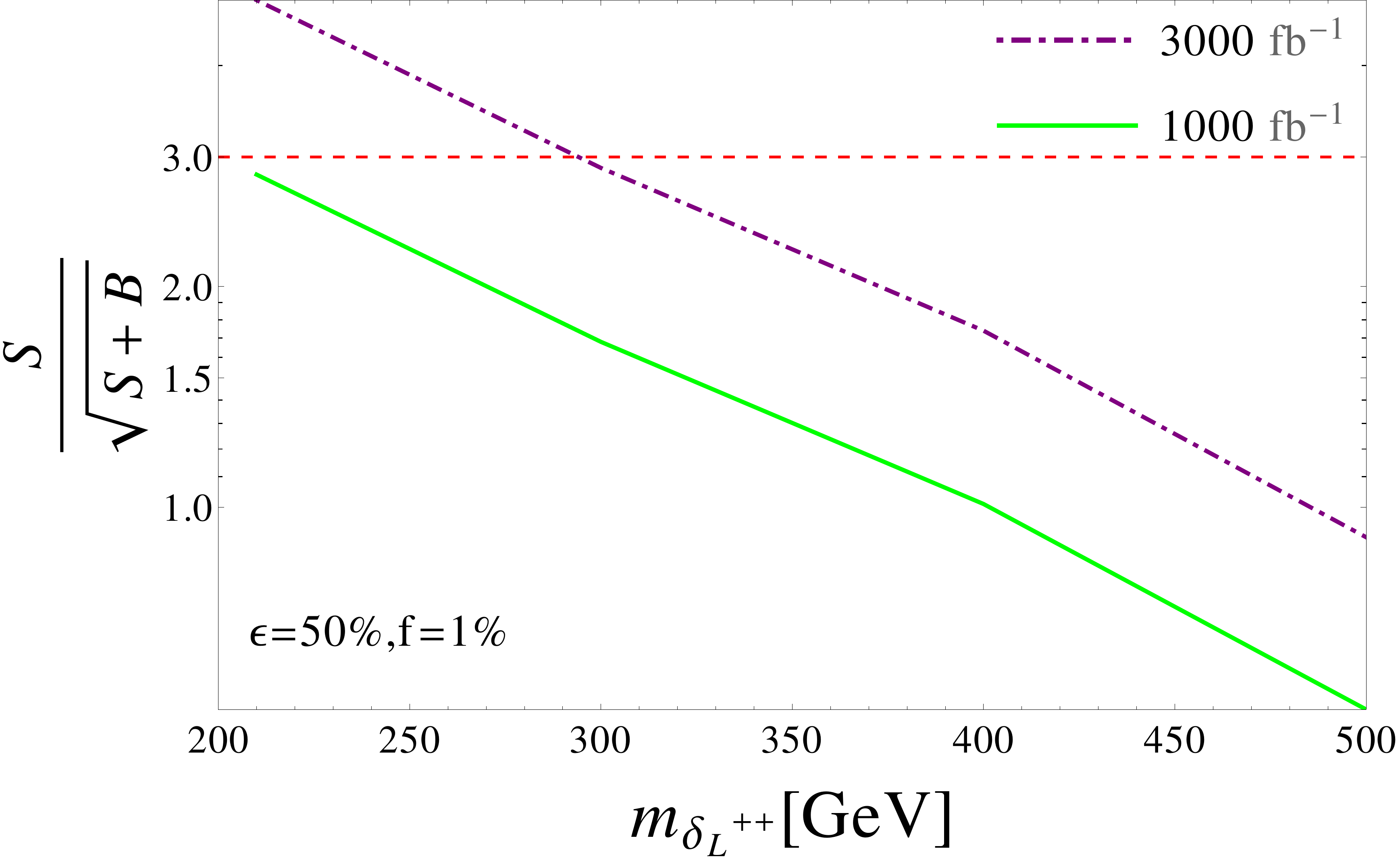}
 \includegraphics[width=3 in]{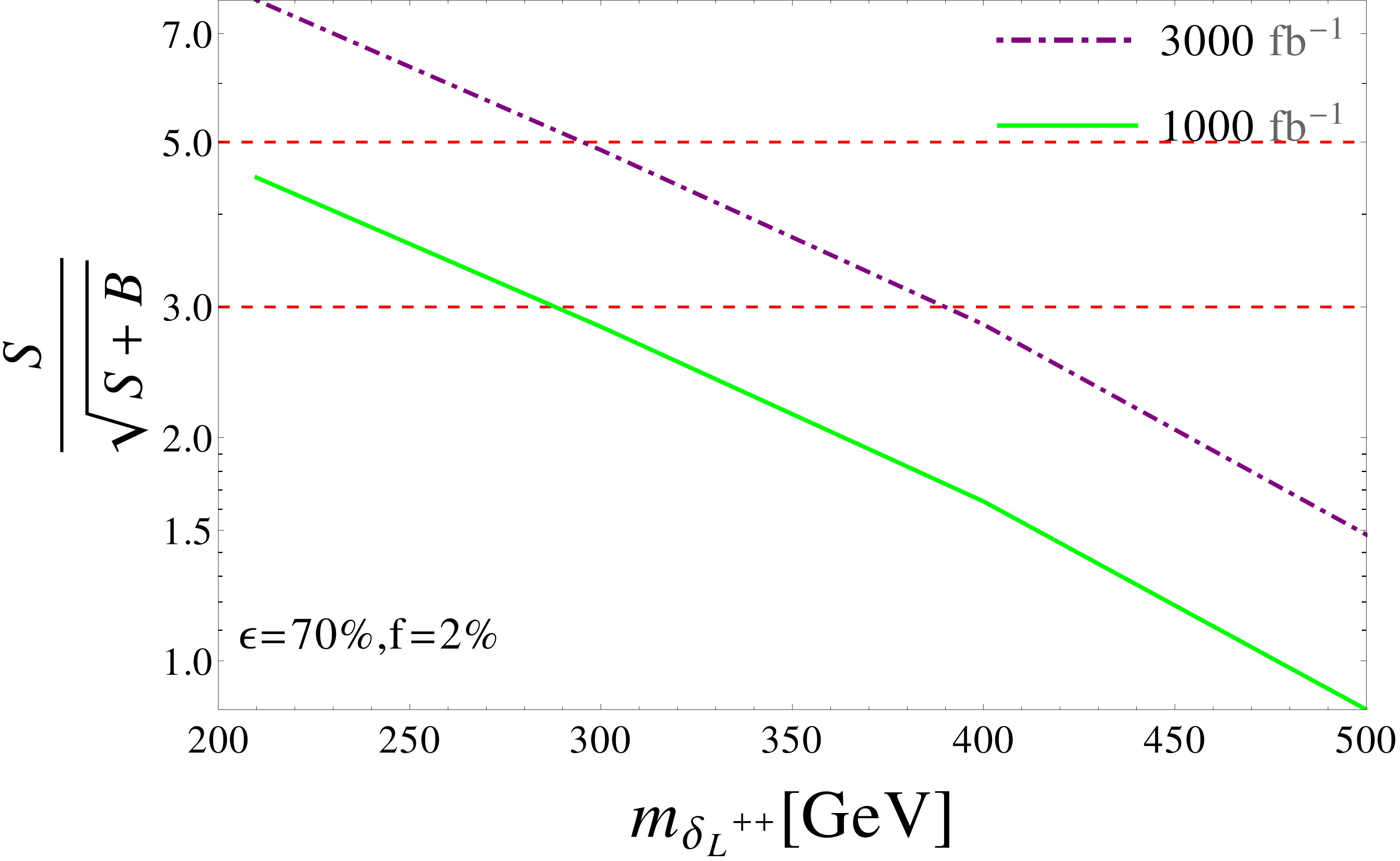}  
\caption{The $\tau_h$ channel significance versus $m_{\delta^{++}_L}$ for $\tau_h$ ID efficiency $\epsilon=50\%$ with the jet fake rate $f=1\%$ [Left panel] and  $\epsilon=70\%$, $f=2\%$ [Right panel]. LHC energy is 14 TeV and $\Delta m$ is kept at $40 \gev$. The significance for luminosities at 1000 and 3000 fb$^{-1}$ are plotted as solid and dot-dashed curves respectively.}
\label{tauSig}
\end{figure}

The signal and background cross-sections are listed in Table~\ref{tauX-sec}. The signal significance is plotted as a function of $m_{\delta^{++}_L}$ in Fig.~\ref{tauSig}, where we keep $\Delta m = m_{\delta^{++}_L}-m_{\delta^{+}_L}= 40 \gev$. For a 3000 fb$^{-1}$ luminosity at 14 TeV, $m_{\delta^{++}_L} $ can be probed upto $ 300 \gev$ at $3 \sigma$ with $\epsilon=50 \%, f=1 \%$. With enhanced tagging rates ($\epsilon = 70 \%$,$f = 2 \%$) the ${\delta^{++}_L}$ mass reach goes up to $390 \gev$ and the corresponding discovery ($5\sigma$) limit of  $m_{\delta^{++}_L}$ is 300 GeV. With 1000 fb$^{-1}$ data, however, the $3 \sigma$ limit is 290 GeV for $\epsilon = 70 \%$ and $f = 2 \%$~\footnote{The $V+$jets production (where, $V=W,Z$), however, could be a considerable source of the background in large jet multiplicity environments at the LHC where a jet to $\tau_h$ fake rate is at a level of $1-2\%$. A full analysis, which takes into account all the fake backgrounds using {\tt{DELPHES 3}}~\cite{deFavereau:2013fsa} as the detector simulator, will be performed in a forthcoming study~\cite{Type-II} motivated by the Type-II seesaw mechanism for neutrino mass generation.}.

\begin{table}[!htp] 
\begin{tabular}{|c| c| c| c| c| c|} 
\hline  
     $(m_{\delta^{++}_L}, m_{\delta^{+}}) $ &   & Selection  & Signal  & ZZjj & WZjj    
     \\
  $[\gev]$ &   & Cuts  & [fb] & [fb] & [fb]   
\\   \hline \hline

  & &  Basic cuts  &  $2.222 \pm 0.009$ & $585.9 \pm 1.4$  & $3513 \pm 8$     
  \\
 & &  VBF cuts &  $0.4655 \pm 0.0040$ &  $39.98 \pm 0.36$ & $211.8 \pm 2.1$
  \\  \cline{2-6}
  
  &  & $\geq$ 3 $\tau_h$'s  & $0.0196 \pm 0.0008$   &  $0.0038 \pm 0.0007$  &  $0.0138 \pm 0.0028$  
  \\
  $(210,170)$ & $\epsilon = 50 \%$, $f = 1 \%$ &   $\tau_h \ p_T$ cuts  &  $0.0147 \pm 0.0007 $ &  $0.0016 \pm 0.0005$  & $0.0070 \pm 0.0021$             
  \\   
 & &  $\met$ cut &  $0.0120 \pm 0.0006 $  & $0.0011 \pm 0.0004$  &  $0.0048 \pm 0.0016$  
 \\  \cline{2-6} 
   
  
 &  &  $\geq$ 3 $\tau_h$'s  & $0.0487 \pm 0.0013$  &  $0.0068 \pm 0.0009$  &  $0.0364 \pm 0.0051$  
  \\
 & $\epsilon = 70 \%$, $f = 2 \%$ &   $\tau_h \ p_T$ cuts  & $0.0356 \pm 0.0006$   & $0.0032 \pm 0.0007$  & $0.0168 \pm 0.0034$             
  \\   
 & &  $\met$ cut & $0.0292 \pm 0.0010$ & $0.0020 \pm 0.0005$  & $0.0112 \pm 0.0027$  
  \\  \hline
\end{tabular}
\caption{Summary of the effective cross-sections and the corresponding statistical errors of both the signal and the backgrounds after each cut is applied in the $\tau$ decay scenario at $\sqrt{s}=14 \tev$.}
\label{tauX-sec}
\end{table}

\section{Conclusion}
\label{Conclusion}

In this paper we investigated the LHC prospects of pair production of  the doubly charged Higgs bosons from  L-R symmetric models via the VBF process. The VBF offers a complementary search strategy to the existing studies on the DY process, and can be important in understanding the electroweak origin of the doubly charged Higgs boson.

We chose the $v_L=0$ scenario to avoid $Z$ boson decay into $\delta^{0}_L \delta^{0*}_L$, which is invisible. Due to the constraint from $\rho_{EW}$, the $\delta^{++}_L, \delta^{+}_L$ mass splitting is less than $\sim 40 \gev$, and $\delta^{++}_L$ dominantly decays into like-sign lepton pairs. The LHC signature is consist of multiple ($\geq3$) leptons, 2 tagging jets in the forward region of the detector and missing energy (in $\tau$ final state only). Due to different identification efficiencies, we studied the light lepton ($e,\mu$) and hadronically decaying $\tau_h$ scenarios separately.

A series of kinematical cuts, led by the VBF-topological cuts, the number of leptons in the final state and $p_T$ cuts, are found to be effective against SM backgrounds. For a luminosity of 1000 fb$^{-1}$, we can set an $3\sigma$ exclusion limit upto $620 \gev$ for the light-lepton final state and $290 \gev$ for $\tau_h$ final state. The latter assumes a $\tau$ ID efficiency at
$70\%$ and a $2\%$ jet fake rate. For $3000$ fb$^{-1}$ luminosity, we achieved an exclusion limit at $m_{\delta^{++}_L}\sim 800 \gev$ and $390 \gev$ in the light leptons ($e,\mu$) and $\tau$ final states.

\section*{Acknowledgements}

 We thank Kechen Wang, Alfredo Gurrola, Sanjay Padhi, Goran Senjanovic, Francisco del Aguila and Olivier Mattelaer for helpful discussions. We would also like to thank Youngdo Oh for reading the manuscript. This work is supported in part by DOE Grant No. DE-FG02-13ER42020, the World Class University (WCU) project through the National Research Foundation (NRF) of Korea funded by the Ministry of Education, Science, and Technology (Grant No. R32-2008-000-20001-0). T.K. is also supported in part by Qatar National Research Fund under project NPRP 5-464-1-080. YG is supported by Mitchell Institute for Fundamental Physics and Astronomy.

\section*{Appendix}
\appendix

\section{The Higgs-sector general potential of the left-right model}
\label{A}

 The potential of the scalar sector of the model we have studied, following Refs.~\cite{Gunion,Desh}, is given by,

\begin{align}
V =-\mu^{2}_1\Tr{\phi^{\dagger}\phi}+\lambda_1(\Tr{\phi^{\dagger}\phi})^2+
  \lambda_2\Tr{\phi^{\dagger}\phi\phi^{\dagger}\phi}+  \dfrac{1}{2}\lambda_3(\Tr{\phi^{\dagger}\tilde{\phi}}+\Tr{\tilde{\phi^{\dagger}}\phi})^2
  +\dfrac{1}{2}\lambda_4(\Tr{\phi^{\dagger}\tilde{\phi}}-\Tr{\tilde{\phi^{\dagger}}\phi})^2
\nn\\ + \lambda_5\Tr{\phi^{\dagger}\phi\tilde{\phi^{\dagger}}\tilde{\phi}}+ \dfrac{1}{2}\lambda_6(\Tr{\phi^{\dagger}\tilde{\phi}\phi^{\dagger}\tilde{\phi}}+
\Tr{\tilde{\phi^{\dagger}}\phi\tilde{\phi^{\dagger}}\phi})  -  \mu^{2}_2(\Tr{\Delta^{\dagger}_L\Delta_L}+Tr{\Delta^{\dagger}_R\Delta_R})
 \nn\\ +\rho_1[(\Tr{\Delta^{\dagger}_L\Delta_L})^2+\Tr{\Delta^{\dagger}_R\Delta_R})^2]
 +\rho_2(\Tr{\Delta^{\dagger}_L\Delta_L\Delta^{\dagger}_L\Delta_L}+
\Tr{\Delta^{\dagger}_R\Delta_R\Delta^{\dagger}_R\Delta_R})
 +\rho_3(\Tr{\Delta^{\dagger}_L\Delta_L})(\Tr{\Delta^{\dagger}_R\Delta_R})
 \nn\\ +\rho_4(\Tr{\Delta_L\Delta_L}\Tr{\Delta^{\dagger}_R\Delta^{\dagger}_R}  
+\Tr{\Delta^{\dagger}_L\Delta^{\dagger}_L}\Tr{\Delta_R\Delta_R})
 +\alpha_1\Tr{\phi^{\dagger}\phi}(\Tr{\Delta^{\dagger}_L\Delta_L}+\Tr{\Delta^{\dagger}_R\Delta_R})
\nn\\  +\alpha_2(\Tr{\Delta^{\dagger}_R\phi^{\dagger}\phi\Delta_R}  
+\Tr{\Delta^{\dagger}_L\phi\phi^{\dagger}\Delta_L})
 +\alpha^{'}_2(\Tr{\Delta^{\dagger}_R\tilde{\phi^{\dagger}}\tilde{\phi}\Delta_R}
+\Tr{\Delta^{\dagger}_L\tilde{\phi}\tilde{\phi^{\dagger}}\Delta_L}),
\end{align}
 where $\tilde{\phi}\equiv\tau_2\phi^{*}\tau_2$. $\tau_i/2$ is the usual $2\times2$ representation matrices of $SU(2)$. One should note that additional discrete symmetries, $\Delta_L \rightarrow \Delta_R$,$\Delta_R \rightarrow -\Delta_R$ and $\phi \rightarrow i \phi$ have been imposed in addition to L-R symmetry, to have $v_L = 0$ at the natural minima of the potential \cite{Gunion}. As it has been already mentioned in Eq.~\eqref{para_def}, we have defined a combination of parameters appeared in the above Higgs potential as:
 \begin{eqnarray}
\rho_{dif}\equiv \rho_3-2(\rho_2+\rho_1),\nn\\ \Delta\alpha\equiv(\alpha_2-\alpha^{'}_2)/2.
\end{eqnarray}

 The mass-spectrum of the left Higgs triplet, obtained after minimising the above potential, have been discussed below. We have neglected terms of order $v_L/v_R$ here.

The left handed doubly charged Higgs boson $\delta^{++}_L$ is a physical mass eigenstate itself with mass\begin{eqnarray}
m^{2}_{\delta^{++}_L}=\dfrac{1}{2}\rho_{dif}v^{2}_R-\rho_2 v^{2}_L+\Delta\alpha\kappa^{2}_1.
\end{eqnarray}

The physical singly charged Higgs boson is
\begin{eqnarray}
\tilde{\delta^{+}_L}=\dfrac{\delta^{+}_L+\dfrac{\sqrt{2} v_L}{\kappa_1} \phi^{+}_2}{(1+\dfrac{2 v^{2}_L}{\kappa^{2}_1})^{1/2}}.
\end{eqnarray} with mass\begin{eqnarray}
m^{2}_{\delta^{+}_L}=\dfrac{1}{2}\rho_{dif}v^{2}_R+\dfrac{1}{2}\Delta\alpha(\kappa^{2}_1+2 v^{2}_L).
\end{eqnarray}

The neutral Higgs sector consists of two pure states $\delta^{0}_L$ and $\delta^{0*}_L$, which have the same mass
\begin{eqnarray}
m^{2}_{\delta^{0}_L}=m^{2}_{\delta^{0*}_L}=\dfrac{1}{2}\rho_{dif}v^{2}_R.
\end{eqnarray}

 The complete set of mass eigenstates of the Higgs sector of the L-R model is discussed in~\cite{Gunion,Desh}. The results presented above are valid for the both vev scenarios under consideration. We can obtain the formulas for relevant cases by simply recalling the fact that for $v_L\neq0$, $\rho_{dif}=0$, while for $v_L=0$, $\rho_{dif}$ can be varied. Thus we can readily observe that in the second scenario $\delta^{+}_L$ is also a pure mass eigenstate.

  \section{The Feynman rules for the left Higgs triplet-gauge interaction}
  \label{C}

  We have presented the Feynman Rules for the interaction of the vector bosons and the left Higgs triplet scalars in Table \ref{FR}. 

\begin{table}[!htp] 
\begin{tabular}{cc}
\hline \hline 
 Vertex & Feynman Rule \\ 
 \hline \hline
$\gamma_{\mu}\gamma_{\nu}\delta^{+}_L\delta^{-}_L$ & $2 i g^2 s^{2}_W g_{\mu\nu}$ \\ 
 
 $\gamma_{\mu}\gamma_{\nu}\delta^{++}_L\delta^{--}_L$ & $8 i g^2 s^{2}_W g_{\mu\nu}$ \\ 
 
 $\gamma_{\mu}\delta^{+}_L\delta^{-}_L$ & $igs_{W}(p_2-p_3)_{\mu}$ \\ 
  
 $\gamma_{\mu}\delta^{++}_L\delta^{--}_L$ & $2igs_{W}(p_2-p_3)_{\mu}$ \\ 
 \hline 
 $\gamma_{\mu}\delta^{0}_L\delta^{-}_L W^{+}_{\nu}$ & $i g^2 s_W g_{\mu\nu}$  \\ 
 
 $\gamma_{\mu}\delta^{+}_L\delta^{--}_L W^{+}_{\nu}$ & $-3 i g^2 s_W g_{\mu\nu}$ \\ 
 
 $\gamma_{\mu}\delta^{-}_L W^{+}_{\nu}$ & $i g^2 s_W v_L g_{\mu\nu}/\sqrt{2}$  \\
 \hline 
 $\delta^{0}_L\delta^{-}_LW^{+}_{\mu}$ & $i g (p_1-p_2)_{\mu}$ \\ 
  
 $\delta^{+}_L\delta^{--}_LW^{+}_{\mu}$ & $-i g (p_1-p_2)_{\mu}$ \\ 
  
  $\delta^{0}_L \delta^{--}_L W^{+}_{\mu} W^{+}_{\nu}$ & $-2 i g^2 g_{\mu\nu}$ \\ 

 $\delta^{--}_L W^{+}_{\mu} W^{+}_{\nu}$ & $-i \sqrt{2} g^2 v_L g_{\mu\nu}$ \\ 
 
 $\delta^{0}_L \delta^{0*}_L W^{+}_{\mu} W^{-}_{\nu}$ & $i g^2 g_{\mu\nu}$ \\ 
  
 $\delta^{+}_L \delta^{-}_L W^{+}_{\mu} W^{-}_{\nu}$ & $2 i g^2 g_{\mu\nu}$ \\ 
  
 $\delta^{++}_L \delta^{--}_L W^{+}_{\mu} W^{-}_{\nu}$ & $i g^2 g_{\mu\nu}$ \\ 

 $\delta^{0}_L W^{+}_{\mu} W^{-}_{\nu}$ & $i g^2 v_L g_{\mu\nu}/\sqrt{2}$  \\
 \hline 
 $\gamma_{\mu}\delta^{++}_L\delta^{--}_L Z_{\nu}$ & $4 i g^2 s_W c_{2W} g_{\mu\nu}/c_W$ \\ 
 
 $\gamma_{\mu}\delta^{+}_L\delta^{-}_L Z_{\nu}$ & $-2 i g^2 s^{3}_W g_{\mu\nu}/c_W$  \\
 \hline 
 $\delta^{0}_L \delta^{-}_L W^{+}_{\mu} Z_{\nu}$ & $-i g^2 (1+s^{2}_W) g_{\mu\nu}/c_W$ \\ 
  
 $\delta^{+}_L \delta^{--}_L W^{+}_{\mu} Z_{\nu}$ & $-i g^2 (1-3 s^{2}_W) g_{\mu\nu}/c_W$ \\ 
 
 $\delta^{-}_L W^{+}_{\mu} Z_{\nu}$ & $-i g^2 v_L (1+s^{2}_W) g_{\mu\nu}/(\sqrt{2}c_W)$  \\
 \hline 
 $\delta^{0}_L \delta^{0*}_L Z_{\mu} Z_{\nu}$  & $2 i g^2  g_{\mu\nu}/c^{2}_W$ \\ 
 
 $\delta^{++}_L \delta^{--}_L Z_{\mu} Z_{\nu}$ & $2 i g^2 c^{2}_{2W} g_{\mu\nu}/c^{2}_W$ \\ 
 
 $\delta^{+}_L \delta^{-}_L Z_{\mu} Z_{\nu}$ & $2 i g^2 s^{4}_{W} g_{\mu\nu}/c^{2}_W$ \\ 
 
 $\delta^{0}_L Z_{\mu} Z_{\nu}$ & $ i \sqrt{2} g^2 v_L g_{\mu\nu}/c^{2}_W$ \\ 
 
 $\delta^{0}_L \delta^{0*}_L Z_{\mu}$ & $- i g  (p_1-p_2)_{\mu}/c_W$ \\ 
  
 $\delta^{+}_L \delta^{-}_L Z_{\mu}$ & $- i g s^{2}_W (p_1-p_2)_{\mu}/c_W$ \\ 
 
 $\delta^{++}_L \delta^{--}_L Z_{\mu}$ & $ i g c_{2W} (p_1-p_2)_{\mu}/c_W$ \\  
  \hline \hline 
 \end{tabular}     
\caption{Feynman Rules for the interaction of the vector bosons and the left Higgs triplet scalars of the L-R Symmetric Model. Here $p_i$ stands for the 4-momentum of the i-th particle at the vertex, with the convention that all the particle momenta are coming into the vertex. For brevity $\cos{2\theta_W}$ has been abbreviated as $c_{2W}$.}
\label{FR}
\end{table}


\end{document}